\documentclass[a4paper,UKenglish,cleveref, autoref, thm-restate]{lipics-v2021}
\usepackage{amsmath}
\usepackage{algorithm2e}
\usepackage{placeins}

\long\def\full#1{%0
{#1}%
}
\long\def\conf#1{%
{}%
}

\newtheorem{invariant}{Invariant}

\title{Succinct Planar Encoding with Minor Operations} %TODO Please add

\author{Frank Kammer}{THM, University of Applied Sciences Mittelhessen, Giessen,
Germany }{frank.kammer@mni.thm.de}{https://orcid.org/0000-0002-2662-3471}{}{}{}
\author{Johannes Meintrup}{THM, University of Applied Sciences Mittelhessen,
Giessen, Germany
}{johannes.meintrup@mni.thm.de}{https://orcid.org/0000-0003-4001-1153}{Funded by
the Deutsche Forschungsgemeinschaft (DFG, German Research Foundation) --
379157101.}{}{}

\authorrunning{F. Kammer and J. Meintrup}

\Copyright{Frank Kammer and Johannes Meintrup}



\category{} %optional, e.g. invited paper

\relatedversion{} %optional, e.g. full version hosted on arXiv, HAL, or other respository/website
\nolinenumbers %uncomment to disable line numbering

\EventEditors{}
\EventNoEds{0}
\EventLongTitle{}
\EventShortTitle{}
\EventAcronym{}
\EventYear{}
\EventDate{}
\EventLocation{}
\EventLogo{}
\SeriesVolume{}
\ArticleNo{}
\begin{document}

\maketitle

\begin{abstract}

Let $G$ be an unlabeled planar and simple $n$-vertex graph. Unlabeled graphs are
graphs where the label-information is either not given or lost during the
construction of data-structures. We present a succinct encoding of $G$ that
provides induced-minor operations, i.e., edge contractions and vertex deletions.
Any sequence of such operations is processed in $O(n)$ time in the word-RAM
model.\ At all times the encoding provides constant time (per element
output) neighborhood
access and degree queries. Optional hash tables extend the encoding with
constant expected time adjacency queries and edge-deletion (thus, all minor
operations are supported) such that any number of edge deletions are computed in
$O(n)$ expected time. Constructing the encoding requires $O(n)$ bits and $O(n)$
time. The encoding requires $\mathcal{H}(n) + o(n)$ bits of space with
$\mathcal{H}(n)$ being the entropy of encoding a planar graph with $n$ vertices.
Our data structure is based on the recent result of Holm et al.~[ESA 2017] who
presented a linear time contraction data structure that allows to maintain
parallel edges and works for labeled graphs, but uses $\Theta(n \log n)$ bits of
space. We combine the techniques used by Holm et al.\ with novel ideas and the
succinct encoding of Blelloch and Farzan~[CPM 2010] for arbitrary separable
graphs. Our result partially answers the question raised by Blelloch and Farzan
whether their encoding can be modified to allow modifications of the graph.
As a simple application of our encoding, we present a linear time outerplanarity testing
algorithm that uses $O(n)$ bits of space.

\end{abstract}

\section{Introduction}
\label{sec:int}

Graphs are used to model systems as entities and relationships between these
entities. Many graphs that arise in real-world application are very large. This
has given rise to an area of research with the aim of reducing the required
{space~\cite{bacic_et_al:LIPIcs.ISAAC.2021.62,BFS,
DBLP:conf/isaac/ElmasryK16, hagerup20,HEEGER202146, DBLP:journals/algorithmica/KammerMS22}.
Practical} examples include large
road-networks~\cite{strasser_et_al:LIPIcs:2020:12947} or social-network
graphs~\cite{Floreskul_Tretyakov_Dumas_2014}. This has spawned research
inquiries into compact representation of graphs, especially those that posses
certain structural properties. The arguably most well-known such structural
property is planarity. A graph is planar if it can be drawn in the plane without
crossings. In this work we consider the problem of maintaining a {succinct}
encoding of a given graph {under edge contractions and vertex deletions,
referred to as {\textit{induced-minor operations}}}. An edge contraction in a
graph $G=(V, E)$ consists of removing an edge $\{u, v\} \in E$ from the graph
and merging its endpoints to a new vertex $x$\full{\ and adding all neighbors of
$u$ and $v$ to the neighborhood of $x$}. Edge contractions are a vital technique
in a multitude of algorithms, prominent examples include computing minimum
cuts~\cite{10.5555/313559.313605}, practical treewidth
computations~\cite{DBLP:journals/jco/Tamaki19} and maximum
matchings~\cite{edmonds_1965}. {At the end of the paper we extend our result
from induced-minor operations to support all {\em{minor operations}}, which in
addition to edge contractions and vertex deletions includes edge deletions.}
\full{There is a famous series of over 20 publications focused on graph minors by
Robertson and Seymour. For the latest paper of this series
see~\cite{DBLP:journals/jct/RobertsonS12}.}%
We work on unlabeled graphs,
meaning that labels are either not given or lost when constructing
our data structure.
%In the next section we present our main result in an intuitive fashion.

\textbf{Related Work.}
For encoding planar graphs without regards to providing fast access operations,
Keeler et al.~\cite{KEELER1995239} showed an $O(n)$ bits representation. For a
compression within the information theoretic lower bound refer to He et
al.~\cite{doi:10.1137/S0097539799359117}. Due to Munro and
Raman~\cite{doi:10.1137/S0097539799364092} there exists an encoding using $O(n)$
bits that allows constant-time queries which has subsequently been improved by
Chiang et al.~\cite{10.5555/365411.365518}. We build on the succinct
representation due to Blelloch and Farzan, which allows encoding arbitrary
separable graphs and subsequently allows constant-time
queries~\cite{10.1007/978-3-642-13509-5_13}, which builds on the work of
Blanford et al.~\cite{10.5555/644108.644219}. Recently it was shown that the
encoding of Blelloch and Farzan can be constructed using $O(n)$ bits in $O(n)$
time~\cite{cloud22}. Our work can be thought of as extending the encoding
of Blelloch and Farzan to allow contraction operations. For edge contractions
in planar graphs without regard to space-usage, Klein and
Mozes~\cite{kleinmozes} presented an algorithm that runs in $O(\log n)$ time per
contraction. Their work is based on techniques by Brodal and
Fagerberg~\cite{Brodal99dynamicrepresentations}\full{, who showed that a bound
out-degree orientation of a planar graph can be maintained under edge insertions
and deletions in $O(\log n)$ time per update}. For edge and vertex deletions
Holm and Rotenberg showed a data structure that provides any number of such
deletions in $O(n)$ time~\cite{holm_et_al:LIPIcs.STACS.2021.42}. The
state-of-the-art by Holm et al.~\cite{holm_et_al:LIPIcs:2017:7875} provides edge
contractions in $O(n)$ time total. Their data structure allows constant time
(per element output) neighborhood and degree queries and maintains
parallel edges that occur due to merges, while also allowing to view the
graph as "simple", i.e., skipping parallel edges when querying the
neighborhoods. Using optional hashing techniques they provide expected constant
time adjacency queries. Their data structure is based on the well-known notion
of $r$-divisions~\cite{FedericksonN87, Goodrich95, KleinNMSSC13}, a technique we use as well.
The data structure of Holm et al.\ uses %$n\log n +
$\Theta(n \log n)$ bits to store mappings
and initially applies graph transformations that 
increase the number of vertices by a constant factor, making it not succinct,
neither for unlabeled nor labeled graphs.
Even assuming these steps could trivially be adapted to use $o(n)$ bits
when encoding unlabeled graphs, they additionally construct a lookup table for
storing small graphs such that each index encoding a graph with $k$ vertices
uses $\mathcal{H}(k) + \Theta(\mathcal{H}(k))$ bits of space, with
$\mathcal{H}(k)$ the entropy of encoding planar graphs with $k$ vertices. They
then encode graphs of at least $n$ total vertices using such indices, i.e., they
use $\mathcal{H}(n)+\Theta(\mathcal{H}(n))$ bits for storing all such small
graph, i.e., their data structure is not succinct.

\textbf{Our result.}
Let $G$ be an unlabeled planar and simple graph with $n$ vertices. We present a
succinct encoding of $G$ that is able to process edge contractions and vertex
deletions in $O(n)$ time for any number of such modifications. At all times,
the data
structure allows constant time (per element output) neighborhood and degree queries.
Using optional hashing techniques, we provide constant expected time adjacency
queries and can process any number of edge deletions in $O(n)$ expected time.
This partially answers a question posed by Blelloch and Farzan if their encoding
for arbitrary separable graphs can be extended to allow graph
modifications~\cite{10.1007/978-3-642-13509-5_13}. Our data structure maintains
the running time of the state-of-the-art solution by Holm et
al.~\cite{holm_et_al:LIPIcs:2017:7875} for labeled planar graphs, while using
significantly less space. 
The data structure of Holm et al. requires the input graph
to be transformed by replacing each vertex with degree larger than a constant
by a cycle-gadget, which increases the
number of vertices by a linear factor\footnote{Outlined in Appendix~A.1 in the full version of
their paper found on arXiv.}.
Such a transformation is not necessary
using our techniques, but
 our data structure only works
for unlabeled graphs, and we are not able to maintain parallel edges that occur
due to contractions, i.e., our graph is at all times simple. In the next
section we present our main result in an intuitive fashion. Our new result is
based on several previous data structures (Section~\ref{sec:pre}) and on
a table-lookup technique (Section~\ref{sec:table}). In
Section~\ref{sec:blelloch}, we describe and extend a succinct
encoding technique due to Blelloch and
Farzan~\cite{10.1007/978-3-642-13509-5_13}. One of our challenges was to extend
mappings used by Blelloch and Farzan to be semi-dynamic. 
Interestingly,
for this we repurpose a graph data structure of Holm et
al.~\cite{holm_et_al:LIPIcs:2017:7875} and use it to construct
dynamic mappings between vertex labels. The dynamic mappings are
described in Section~\ref{sec:dynmapping}, which
% we
%introduce data structures for several dynamic mappings that 
we use in Section~\ref{sec:result} to combine the results of all the
previous sections to achieve our dynamic encoding.
We end this section
by extending our encoding to provide vertex and edge deletions as
well as adjacency and degree queries. Finally, in Section~\ref{sec:outer},
we present 
a simple application of our encoding, which is outerplanarity testing
of a given graph in linear time using $O(n)$ bits of space, with $n$
the number of vertices of the graph.
\conf{All proofs can be found in the full version of this paper.}
% In Appendix~\ref{app:appli} we present
% two simple applications to demonstrate the use of our encoding in the design of
% space-efficient algorithms: first, we present an implementation of a linear time
% and bit outerplanarity test, which improves on the previous best linear bit
% algorithm of Kammer et al.~\cite{KammerKL19} which has a runtime of $O(n \log
% \log n)$. Secondly, we present the first linear bit and time implementation of a
% maximal planarity test for a graph not knowing it is planar.

\section{A succinct graph encoding for edge contractions}
\label{sec:overview}
We now give an overview of our new data structure for succinctly encoding a
simple unlabeled planar graph $G=(V, E)$ {with $n$ vertices} and
maintaining this encoding under edge contractions while allowing neighborhood
and degree queries in constant time per element output. Vertex deletions and
edge deletions are discussed in Section~\ref{sec:result}. 
Note that while we work with unlabeled graphs, internally we assume vertices are labeled
as consecutive integers from $1$ to $n$. The intuitive idea is
to construct a set $X \subset V$ of \textit{boundary vertices} such that $G[V
\setminus X]$ (the vertex-induced graph on $V \setminus X$) contains multiple
connected components $C_1, \ldots, C_k$ of "small" size at most $r$ (which is
defined later) with $k=O(n / r)$, and $X$ of size $O(n / \sqrt{r})$. Based on
this, we distinguish edges of three types: edges between vertices of $X$ (called
\textit{boundary edges}) edges between vertices of one $C_i$ (called
\textit{simple edges}) and edges between a vertex of $X$ and a vertex of some
$C_i$ (called \textit{mixed edges}). For each $C_i$ denote with $P_i$ the graph
induced by all simple edges between vertices of $C_i$ and all mixed edges with
one endpoint in $C_i$. The set of all these $P_i$ is known as an
\textit{$r$-division}, and each $P_i$ is called a \textit{piece}. Note
that an $r$-division has some additional characteristics which are defined
more precisely later, one such key characteristics that each $P_i$ contains only
$O(\sqrt{r})$ boundary vertices, and therefore is of size
$O(|C_i|+\sqrt{r})=O(r)$. Note that each boundary vertex is contained
in multiple pieces. Assume that there is a
data structure that is able to easily contract any number of edges in graphs of
size $O(r)$ in time $O(r)$. As long as contractions are only carried out between
simple edges, we would be able to easily construct a data structure that results
in $O(n)$ runtime for any number of contractions by simply constructing this
data structure for each piece individually. Problems occur when contracting
boundary or mixed edges because contractions are no longer able to be carried
out locally in a single piece as they affect vertices in multiple pieces. For
example, when contracting a boundary edge $\{u, v\}$ this affects every $P_i$
with $i \in \{1, \ldots, k\}$ that contains at least one of $u$ and $v$. To provide such
contractions we construct a data structure sketched in the following. Note that
we distinguish between vertex merges of two vertices $u, v$ and edge
contractions between an edge $\{u, v\}$, with the latter being analogous to the
first with the distinction that $u$ must be adjacent to $v$.

For edges between vertices of the boundary $X$ we construct a \textit{boundary
graph} $F=G[X]$, which at all times contains all boundary edges, including edges
that occur due to contractions. A key invariant that we uphold is that the
"status" of a vertex, i.e., if it is a boundary or a non-boundary vertex, never
changes due to contractions. If a vertex is a boundary vertex initially,
it will be handled internally as a boundary vertex even when
it no longer is incident to boundary edges due to contractions.
For non-boundary vertices we ensure that these vertices are
never incident to boundary edges.
For each $u \in X$ we maintain a mapping
containing all $i$ with $u \in V(P_i)$, i.e., the pieces that contains $u$. Now,
when we contract a boundary edge $\{u, v\}$ we firstly merge $v$ to $u$ in all
$P_i$ that contain both $u$ and $v$. For any $u \in V$ we denote
with $N(u)$ the neighborhood of $u$. We process the aforementioned merge by
setting $N(u):=(N(u) \cup N(v)) \setminus \{u, v\}$ in $P_i$ and removing $v$
from $P_i$. In all $P_i$
that contain only $v$, we simply relabel $v$ to be $u$. In all $P_i$ that do not
contain $v$ (but can contain $u$) we do not have to make modifications. Finally,
$\{u, v\}$ is contracted in the boundary graph $F$ as well. {To maintain $F$ we
can use a "slow" data structure, as $F$ is small.} 

To contract a mixed edge $\{u, v\}$ with $u \in X$ we know there is only one
$P_i$ that contains $v$, where we execute the contraction. This might result in
$u$ now being adjacent to some $x \in X \cap V(P_i)$, for which we add the
respective edge $\{u, x\}$ to $F$. It helps to achieve our runtime goal of $O(n)$ for
processing any number of edge contractions if we do not add this edge
$\{u, x\}$ to all other pieces that contain both $u$ and $x$. We therefore
maintain a second invariant: that edges between boundary vertices are only
contained in the boundary graph (for now ignoring the specifics of how this is
maintained).

 To handle contractions inside pieces, intuitively we build the same data
structure we outlined here to one more
time, splitting each piece in tiny pieces of size at most $r'$, specified later.
We can categorize all graphs of size at most $r'$ by a lookup table, which
allows us to encode every such tiny piece as an index into the lookup table
(Section~\ref{sec:table}). For each
graph of the lookup table we pre-compute all possible vertex merges. We then
contract edges in constant time by simply retrieving (the index of) the
contracted graph from the lookup table. Such a framework was previously used by
Holm et al.~\cite{holm_et_al:LIPIcs:2017:7875} to maintain a planar graph under
contractions, but with a space usage of $\Theta(n \log n)$ bits.

To output the neighborhood of a vertex we distinguish between two cases. First,
to output the neighborhood of a vertex $u$ that is not a boundary vertex, we can
simply output the neighborhood in the only $P_i$ that contains $u$. For a
boundary vertex $u$ we first output the neighborhood of $u$ in $F$, which are
all neighbors being boundary vertices, and then do the same for each piece $P_i$
that contains neighbors of $u$.
%To summarize, the
%boundary graph $F$ maintains edges between boundary vertices. 
%Moreover, each piece $P_i$ maintains simple and mixed edges. The mapping for a
%vertex $u$ of $F$ tells us in which $P_i$ it is part of at least one mixed edge.

To achieve a succinct data structure we build on a succinct encoding \full{for
planar and other separable graphs described by}\conf{due to} Blelloch and
Farzan~\cite{10.1007/978-3-642-13509-5_13}, outlined in
Section~\ref{sec:blelloch}.\full{\ They also construct graph divisions in a
two-level fashion as an encoding scheme. In detail, they}\conf{\ They} construct
an $r$-division for the input graph, and for each piece $P_i$ of the
$r$-division construct another $r'$-division. Each piece $P_{i, j}$ of this
$r'$-division is categorized by a succinct index in a lookup table\full{,
referencing an isomorphic graph $P^*_{i, j}$ which uses "local" labels from $1
\ldots r'$, instead of a subset of the "global" labels $1 \ldots n$}. They
use succinct dictionaries to translate information between the two levels.
These translation data structures and the lookup table can be realized
with $o(n)$ bits. \full{The majority of the space to store the encoding is taken up by
storing one index into the lookup table for each graph of size $O(r')$.} We use
the encoding of Blelloch and Farzan, but enhance it with dynamic qualities
(Section~\ref{sec:blelloch} and Section~\ref{sec:dynmapping}). This (partially)
answers the question posed by Blelloch and
Farzan~\cite{10.1007/978-3-642-13509-5_13}, if modifications of the succinctly
encoded graph are possible. A key notion is that most of the novel "building
blocks" handle the small number of boundary vertices, so non-space efficient data structures are used.

\newcommand{\thmmain}{
\begin{theorem}
    \label{thm:main}
    Let $\mathcal{H}(n)$ be the entropy of encoding a planar graph with $n$
    vertices and $G$ an unlabeled simple $n$-vertex planar graph. There exists
    an encoding of $G$ that provides induced-minor operations (i.e., vertex
    deletions and edge contractions) with the following properties: The encoding
    requires $O(n)$ time to execute any number of {induced-minor
    operations} and provides neighborhood and degree operations in constant time
    (per element output). The encoding requires $\mathcal{H}(n)+o(n)$ bits and
    initialization can be done in $O(n)$ time with $O(n)$ bits.
\end{theorem}
}
\thmmain

Using optional hashing techniques adapted from the data structure of Holm et al.
\cite[Lemma~5.15]{holm_et_al:LIPIcs:2017:7875} we provide edge deletion and
adjacency queries.

\newcounter{countlastcor}
\setcounter{countlastcor}{\thetheorem}

\newcommand{\coredge}{
\begin{corollary}
    \label{cor:edge}
    The encoding of Theorem~\ref{thm:main} can be extended to provide expected
    $O(1)$ time adjacency queries and process any number of minor operations in
    expected $O(n)$ time.
\end{corollary}
}
\coredge

\section{Preliminaries}
\label{sec:pre}
We denote by $[k]=\{1, \ldots, k\}$, with $k$ any integer. Our model of
computation is the word-RAM with a word length of $w = \Omega(\log n)$ bits. We
work with\conf{\ simple} unlabeled graphs\full{,\ which means that we are free
to use an arbitrary labeling for our final encoding. For constructing our
succinct encoding using $O(n)$ time and bits we require the input graph to be
represented by adjacency arrays instead standard adjacency lists, i.e., random
access queries are required. If one only requires a bound of $\Theta(n \log n)$
during the construction phase, this stipulation can be dropped}. In the following
let $G=(V, E)$. We use $G[V']$ with $V'\subseteq V$ to denote the vertex induced
subgraph on $V'$ of $G$. We also write $V(G)$ for the vertices $V$ of $G$ and
$E(G)$ for~$E$. A
merge of a pair of two vertices $u, v \in V$ means replacing both vertices $u,
v$ with a vertex $x$ with $N(x) = N(u) \cup N(v) \setminus \{u, v\}$. In our
data structures we merge $u$ and $v$ by setting $N(u):=N(u) \cup N(v) \setminus
\{u, v\}$ and removing $v$ from $V$, i.e., $x$ is either $u$ or $v$.
We then say that \textit{$v$ is merged to
$u$}. Note that merging $u$ to $v$ is a different operation. Merging two
adjacent vertices $u, v$ is called \textit{contracting the edge $\{u, v\}$}. We
denote by $n$ the number of vertices of the graph $G$ under consideration. 
In this work we use $n_G$ to denote the number of vertices of a given graph $G$.
If the graph is clear from the context we simply write $n$.%If
%there are multiple graphs in a context we use subscript notation to distinguish
%between graphs, i.e., when talking about a graph $G_i$ for some integer $i$ we
%use $n_i$ as the number of its vertices.

\textbf{Planar graph.} A graph is planar exactly if it can be drawn in the plane
such that no two edges cross. The family of planar graphs is closed under taking
minors. Any simple planar graph $G$ has $O(n)$ edges. We only work with simple
graphs and from this point onward assume all (planar) graphs are simple. An operation
that modifies a planar graph under consideration is \textit{planarity
preserving} if afterwards the graph is still planar. For brevity's sake, we
assume all merges discussed in our work are planarity preserving unless stated
otherwise. We denote with $\mathcal{H}(\cdot)$ the entropy to encode a planar
graph, a function dependent on the number of vertices of the graph under
consideration. It is known that
$\mathcal{H}(n)=\Theta(n)$~\cite{DBLP:journals/dam/Turan84}.
\full{Bonichon et al. showed an upperbound for $\mathcal{H}(n)$ of
roughly $5.07n+O(\log n)$ bits, with a lowerbound of roughly $4.71n+O(\log
n)$~\cite{BonichonGH03}.}

We assume\ w.l.o.g.\ that all input graphs for our data structure
are connected. If this is not the case, we add a dummy vertex and connect it to
an arbitrary vertex in each connected component. After our encoding of
Theorem~\ref{thm:main} is constructed, we can simply ignore the dummy vertex and
all its incident edges during any output of the provided operations. As we
require this modification to be space-efficient, concretely that it only uses
$O(n)$ bits of additional space during the construction and runs in $O(n)$ time,
we provide an explicit lemma for this modification. Note that a succinct
encoding of a planar graph with this additional dummy vertex requires only a
constant number of additional bits, as $\mathcal{H}(n+1)=\mathcal{H}(n)+O(1)$.

\newcounter{countlemconnected}
\setcounter{countlemconnected}{\thetheorem}
\begin{lemma}
    \label{lem:connected}
    Let $G$ be a simple planar graph. We can add a dummy vertex $v_d$ to $V$ and
    all edges $\{v_d, u\}$ to $E$ with $u$ being one vertex in each connected
    component of $G$ using $O(n)$ bits in $O(n)$ time.
\end{lemma}
\full{
\begin{proof}
    Identify an arbitrary vertex in each connected component using a
    breadth-first search (BFS). {Denote with $S$ the set of all these
    vertices. This set is exactly the neighborhood of the dummy vertex $v_d$ we
    add to $G$, i.e., $N(v_d)=S$.} We represent the set $S$ with a bit-array of
    length $n$ with a bit set to $1$ at an index $u$ exactly if the vertex $u$
    is in $S$. Using a space-efficient BFS that runs in $O(n)$ time and uses
    $O(n)$ bits~\cite{BFS} we arrive at our desired runtime and space-usage. It
    is easy to see that this modification preserves planarity. {To
    support standard access operations, we construct an indexable dictionary
    (Lemma~\ref{lem:ID}) for the set $S$, which uses $O(n)$ bits total. This
    allows us to iterate over the neighborhood of $v_d$ in constant time per
    element. When iterating over the neighborhood of any vertex $u \in V
    \setminus \{v_d\}$ first check if $S[u]=1$ (in the bit-vector representing
    $S$), in which case output $v_d$ as the first neighbhor of $u$. Afterwards,
    continue with the neighborhood iteration in $G$ normally, i.e., output all
    other neighbors of $u$.}
\end{proof}
}

\textbf{Graph divisions.} Let $G=(V, E)$ be a planar graph and $r$ some integer.
An \textit{$r$-division} $\mathcal{R}=\{P_1, \ldots, P_k\}$ of $G$ is a division
of $G$ into $k=O(n/r)$ edge-disjoint connected subgraphs called \textit{pieces}.
Each piece has $O(r)$ vertices. For each $P \in \mathcal{R}$ there exists a set
of \textit{boundary vertices} $\delta P \subset V(P)$ such that $u \in \delta P$
if and only if $u$ is incident to an edge $\{u, v\} \in E$ with $v \notin V(P)$.
For each $P$ it holds that $|\delta P|=O(\sqrt{r})$. We denote with $\delta
\mathcal{R} = \bigcup_{P \in \mathcal{R}} \delta P$ the set of boundary vertices
of $\mathcal{R}$. For any $r$-division $\mathcal{R}$ we denote by $k$ the number
of pieces, and assume they are numbered from $[k]$ as $\mathcal{R}={P_1, \ldots,
P_k}$. We use a subscript numbering to distinguish between multiple
$r$-divisions as follows: we use the same subscript numbering to refer to the
number of parts of the $r$-division, i.e., we use $k_i$ when talking about an
$r$-division $\mathcal{R}_i$ for some integer $i$. Linear time
algorithms for computing $r$-divisions exists~\cite{Goodrich95, KleinNMSSC13}.
Note that an $r$-division requires each piece to be connected, and our encoding
in some sense maintains a dynamic $r$-division. Due to modifications, pieces may
become disconnected. We do not require each piece to be connected once the
encoding is constructed, this abuses the definition of $r$-divisions without
consequence.

% \begin{lemma}[\cite{10.1007/978-3-642-13509-5_13}]
%     Let $G$ be a planar connected graph and $\mathcal{S}=\{G_1, \ldots, G_k\}$ a
%     separator hierarchy of $G$ with $k=O(n / r)$ and $|V(G_i)| < r$ for all $i
%     \in [k]$ and $r$ some fixed value. There are $O(n / \sqrt{r})$ duplicate
%     vertices and each $G_i$ has $O(\sqrt{{r}})$ boundary vertices.
% \end{lemma}

\textbf{Forbidden-vertex graph data structure.} 
We use a so-called forbidden-vertex graph data structure that is 
initialized for a simple planar graph $G=(V, E)$
and any set $B \subseteq V$ of \textit{forbidden vertices}. It allows
modifications of $G$ in the form of edge insertions and deletions, and in the
form of vertex merges while maintaining two invariants: no edges between
vertices of $B$ exist and $G$ is simple and planar. This data structure was
described by Holm et al.~\cite{holm_et_al:LIPIcs:2017:7875} as a building block
for their contraction data structure. We slightly modify their data structure
and change some notation to match our use-cases. We use this data
structure (among other things) for maintaining edges between so-called boundary
vertices, as sketched in Section~\ref{sec:overview}.
 We refer to this data structure as
\textit{forbidden-vertex graph data structure}. 

For each vertex and edge managed
by the data structure we can access and modify auxiliary data, which takes
constant time per word written or read, if a reference to the vertex or edge is given. When
merging two vertices $u, v$ some edge $\{u, x\}$ might be removed and
inserted again as $\{v, x\}$. We view this as the same edge, but with different
endpoints. Meaning, auxiliary data of $\{u, x\}$ is now stored at $\{v, x\}$. If
$\{v, x\}$ already existed before the merge, we can decide what to do with the
data of the discarded parallel edge. Self-loops and forbidden edges that would occur
due to a merge
are output during the merge operation. %The data structure also provides adjacency,
%neighborhood and degree queries in constant time (per element).
% Unsere Graphen haben keine self-loops, also können Kanten, die zum
% self-loop werden, nicht mehr wieder auftauchen. 
%Analogous for occurring self-loops. 
All operations are only permitted if they preserve
planarity. In the following we more precisely define the available
modifications:

\begin{itemize}
    \item \textbf{Merge.} Given are two vertices $u, v$ with $u \neq v$. Merge
    $v$ to $u$ by setting $N(u) := N(u) \cup N(v) \setminus \{u, v\}$ and
    removing $v$ and all incident edges from $V$. Returns a reference to $u$ and
    reports and discards all parallel edges during the merge, and reports all
    non-parallel edges inserted to $N(u)$ during the merge. Edges that would
    occur between vertices of $B$ are discarded.

    \item \textbf{Insert.} Given are two vertices $u, v$ with $u \neq v$ and
    $\{u, v\} \notin E$. Insert the edge $e=\{u, v\}$ into $E$, unless both $u, v \in B$.

    \item \textbf{Delete.} Let $\{u, v\}$ be a given edge. Remove the edge $\{u,
    v\}$ from $E$.
\end{itemize}

\begin{lemma}[\cite{holm_et_al:LIPIcs:2017:7875}]
    \label{lem:bvm}
    Let $G=(V, E)$ be a simple planar graph and $B\subseteq V$. A forbidden
    vertex graph data structure can be initialized for $G$ and $B$ in $O(n \log
    n)$ time. It provides constant time (per element output) neighborhood and adjacency
    queries and access to the label mappings. Edge insertion/deletion
    takes $O(\log n)$ time. Any number of free-assignment vertex merges are
    executed in $O(n \log^2 n)$ time. The data structure uses $O(n \log n)$
    bits.
\end{lemma}

To achieve the runtime outlined in Lemma~\ref{lem:bvm} for vertex merges, each
merge of two vertices $u, v \in V$ is processed by merging the vertex with the
lowest degree to the vertex with the highest degree,\ i.e., we can not
freely choose which vertex is merged. A simple mapping using standard data
structures allows us to label the vertex $x \in \{u, v\}$ that remains after the
merge either $u$ or $v$. The details of this are found in full version of our paper.
We refer to a merge of two vertices where we are able
to freely decide the labeling of the remaining vertex after the merge as
\textit{free-assignment merge}. We henceforth assume that all merges of the data
structure are free assignment merges. The
following lemma summarizes this.

\newcounter{countlemfreeassignmentmerge}
\newcommand{\lemfreeassignmentmerge}{
  \setcounter{countlemfreeassignmentmerge}{\thetheorem}
\begin{lemma}
    \label{lem:freeassignmentmerge}
    Let $G=(V, E)$ be a simple planar graph and $B\subseteq V$ managed by the
    forbidden-vertex graph data structure of Lemma~\ref{lem:bvm}. Using $O(n)$
    additional time for initialization and $O(n \log n)$ bits we are able to
    process any number of free-assignment merges in $O(n \log^2 n)$ time on $G$.
\end{lemma}
}
\lemfreeassignmentmerge
\newcommand{\prooffreeassignmentmerge}{
\begin{proof}
    Let $G_0=(V_0, E_0)$ be the graph managed by the data structure before any
    contractions, edge insertions or deletions are processed. Construct a
    mapping $\texttt{internal}: V_0 \rightarrow V$ and a mapping
    $\texttt{external}: V \rightarrow V_0$, both initially equal to the identity
    function. Assume we merge two vertices $u$ and $v$ and denote with $x \in
    \{u, v\}$ the vertex that remains after the merge. Denote with $y \in \{u,
    v\}$ the label we want to assign to $x$, i.e., the labeling that turns the
    merge into a free assignment merge. For this, set $\texttt{internal}(y):=x$
    and $\texttt{external}(x):=y$. Given a user-assigned label $u$ of a vertex,
    find the internal labeling via $\texttt{internal}(u)$. All internal labels
    are translated via the $\texttt{external}$ mapping before output during
    any of the operations provided by the simple dynamic graph data structure.
    Both mappings can be implemented with a simple array of length $|V_0|$ in
    time $O(n)$ using $O(n \log n)$ bits. 
\end{proof}
}
\full{\prooffreeassignmentmerge}

\textbf{Indexable dictionaries.}
% We make use of the following lemmas to realize various data structures in later
% sections.
% \begin{lemma}[\cite{hagerup20}, Lemma 5.2]
%     \label{lem:hagerup1}
%     For all integers $n$, $N$, following an $O(n)$-time initialization, an array
% of $n$ initially empty binary strings $s_1,\ldots, s_n$ that at all times
% satisfy $|s_i| = O(\log n)$ for $i \in [n]$ and $\sum_{i=1}^n|s_i| \leq N$ can be
% maintained in $O(n \log \log n + N)$ bits under constant-time reading and amortized
% constant-time writing of individual array entries.
% \end{lemma}
% \begin{lemma}[\cite{hagerup20}, Theorem 5.3]
%     \label{lem:hagerup2}
%     For all integers $n$, $N$, following an $O(n)$-time initialization, an array
%     of $n$ initially empty binary strings $s_1,\ldots, s_n$ that at all times
%     satisfy $|s_i| = O(\log n/\log \log n)$ for $i \in [n]$ and
%     $\sum_{i=1}^n|s_i| \leq N$ can be maintained in $O(n + N)$ bits under
%     constant-time reading and amortized constant-time writing of individual
%     array entries. 
% \end{lemma}
{We use a data structure called \textit{indexable dictionary} (ID),
initialized for a universe $U$ of consecutive integers and a set
$S \subseteq U$ and supports membership, rank and select queries.} A rank query
for some $x \in U$ returns $|\{y \in S: y < x\}|$. A select query for some
integer $i$ returns the value of $x \in U$ such that $x$ is stored at rank $i$.
%An ID can intuitively be thought of as a bit-vector of length $|U|$ with a bit
%set to $1$ at index $x$ exactly if $x \in S$, with the addition that it
%supports the specified rank/select queries. 

\begin{lemma}[\cite{10.1145/1290672.1290680}]
    \label{lem:ID}
    Let $s \leq u$ be two integers. Given a set $S$ of size $s$, which is a
    subset of a universe $U=[u]$, there is a succinct indexable dictionary (ID)
    on $S$ that requires $\log{u \choose s} + o(s) + O(\log \log u)$ bits and
    supports rank/select on elements of S in constant time.
\end{lemma}
We use IDs with $u=n$ and $s=O(n/\Lambda(n))$ for some function
$\Lambda(n)=\omega(1)$, then each ID requires $o(n)$ bits\full{, a characteristic
already used by Blelloch and Farzan~\cite{10.1007/978-3-642-13509-5_13}}.
IDs can be can be constructed in $O(u)$ time using $O(u)$ bits~\cite{10.1145/1290672.1290680}.

\newcommand{\tableops}{
    \begin{itemize}
        \item Range filtered neighborhood iteration: Takes as input an index $i$
        into the lookup table, three vertices $u, a, b \in V$, with $G=(V, E)$ being
        the graph encoded at index $i$ and $u \neq a \neq b$. Provides an iterator
        over all neighbors $v$ of $u$ with $a \leq v \leq b$.
    
        \item Batch edge deletion: Takes as input an index $i$ into the lookup
        table, three vertices $u,a, b \in V$ with $G=(V, E)$ the graph encoded at
        index $i$. Returns the index $j$ encoding the graph $G'=(V, E \setminus X)$
        with $X=\{\{u, v\}: v \in N(u) \text{ and } a \leq v \leq b\}$.
        % only permitted if it preserves planarity.
    
        \item Label-preserving merge: Takes as input an index $i$ into the
        lookup table and two vertices $u, v \in V$, with $G=(V, E)$ being the
        graph encoded at index $i$. Returns the index $j$ encoding the graph
        $G'$ obtained from $G$ by setting $N(u) := N(u) \cup N(v) \setminus \{u,
        v\}$ and deleting $v$\full{\ (see the next paragraph for
        details on this)}. Only allowed when preserving planarity.
    \end{itemize}
}

\newcounter{countlemtable}
\newcommand{\lemtable}{
  \setcounter{countlemtable}{\thetheorem}
  \begin{lemma}
    \label{lem:table}
    Let $\ell$ be a positive integer. There exists a table that encodes all planar
    graphs with vertex set $\{1, \ldots, \ell'\}$ for all integers $\ell' \leq \ell$ with
    the following properties. For every graph encoded by the table, (range
    filtered) neighborhood iteration, adjacency queries and label-preserving
    merge operations and batch edge deletion are provided in constant time (per
    element). The table can be constructed in $O(2^{\mathrm{poly}(\ell)})$ time
    using $O(2^{\mathrm{poly}(\ell)})$ bits. %The table contains
    %$2^{\texttt{poly}(r)}$ entries with $\mathcal{H}(r)$ the entropy of encoding a
    %planar graph with $r$ vertices. 
    Every index of the table referencing a graph with $\ell'$ vertices requires
    $\mathcal{H}(\ell')+o(\ell')$ bits.
\end{lemma}

}

\newcommand{\sectable}{
\section{Table lookup for small planar graphs}
\label{sec:table}
In this section we present our table lookup data structure for small graphs.
Given an integer~$\ell$ the table lists for every positive integer~$\ell' \leq \ell$
every possible planar graph with at most $\ell'$ vertices. Such a lookup table was
used by Blelloch and Farzan~\cite{10.1007/978-3-642-13509-5_13} as a building
block for succinctly encoding planar and other separable graphs, which we
outline Section~\ref{sec:blelloch}. For every graph $G$ encoded by the table,
they provide adjacency queries and neighborhood iteration in constant time (per
element). The table can be realized using $O(2^{\texttt{poly}(\ell)})$ bits and
time, including the data structures needed to provide the queries. To
distinguish between all planar graphs with $\ell'$ vertices we requires $\mathcal{H}(\ell')+O(1)$
bits. This corresponds to an index into the computed table. The table
contains $2^{\mathcal{H}(\ell)}$ entries. Everything mentioned so far was shown by
Blelloch and Farzan. In the following we introduce additional operations
and extensions of this lookup table. These modifications increase
the size of the table by a negligible amount of bits while maintaining
the same (asymptotical) runtime for constructing the table. We show that
our modifications increase the size of each index encoding a graph with $\ell'$
vertices by $o(\ell')$ bits, which is negligible for our use case.

\tableops

All these operations can easily be pre-computed for one entry of the table in
time $O({\mathrm{poly}(\ell)})$ using the same amount of bits. The sum of
computations over every entry of the table is $O(2^{\mathrm{poly}(\ell)})$. As a
note on the label-preserving merge operation, vis-à-vis keeping the vertex $v$,
but marking it deleted, consider the following example. We want to merge the
vertices $u$ and $v$ in some graph $G=(V, E)$ encoded by the table with $\ell'$
vertices. If we would simply merge them, the vertex $v$ no longer exists. In
particular, the vertex set after the merge is $V'=V \setminus \{v\}$. As the
vertex set for each graph encoded by the table is consecutively numbered from
$[\ell']$, graphs with vertex set $V'$ are possibly not encoded by the table. To
remedy this, we simply keep the vertex $v$ in the graph but mark it deleted. It
remains to show how to handle 'mark $v$ as deleted'. To each graph encoded by
the table we add a dummy vertex called $\texttt{deleted}$. The table lists every
possible planar graph with at most $\ell+1$ vertices, where the vertex with the
largest label is our dummy vertex $\texttt{deleted}$. To mark $v$ as deleted
during the label preserving merge, we set
$N(\texttt{deleted}):=N(\texttt{deleted}) \cup \{v\}$ (and $N(v):=
\{\texttt{deleted}\}$). We are now able to check if a vertex is deleted, by
checking if it is adjacent to $\texttt{deleted}$. When we later encode a graph
$G=(V, E)$ via an index into this lookup table, we encode it as the graph
$G'=(V\cup \{\texttt{deleted}\}, E)$, i.e., with no vertices marked as deleted
initially. 

It remains to observe how one additional extra vertex increases the size of the
table. The size of each entry encoded by the table stays asymptotically the
same, and is thus of no concern to us. The number of entries in the table is
$O(2^{\mathcal{H}(\ell+1)})$, and therefore each index into the lookup table
encoding a graph with $\ell'$ vertices requires $\mathcal{H}(\ell'+1)+O(1)$ bits to be
stored. As $\mathcal{H}(\ell'+1)=\Theta(\ell'+1)$~\cite{DBLP:journals/dam/Turan84} it
holds that $\mathcal{H}(\ell'+1)=\mathcal{H}(\ell')+O(1)$, which is negligible for our
purpose. Therefore, our table uses $2^{\texttt{poly}(\ell)}$ bits, which is
asymptotically the same as the original table of Blelloch and Farzan. Each index
of the table uses only a constant number of additional bits over the theoretical
lower bound. Note that indices into the original unmodified table of Blelloch
and Farzan also require this additional $O(1)$ bits. We introduce some further
modifications that increase the additional space per index storing
a graph with $r'$ vertices by $o(r')$ bits, which is fine for our use-case.

    When we later use the table lookup to contract edges, we do so by
    effectively replacing an unlabeled graph with a different unlabeled one.
    Without care, this can break internal labeling structures, e.g., a vertex in
    a graph encoded by the lookup table has an internal label of $5$, and after
    replacing the graph it now has a label of $7$. Section~\ref{sec:dynmapping}
    and Section~\ref{sec:result} show how additionally maintain a dynamic label
    mapping structure for "important" vertices, i.e., boundary vertices as
    described in Section~\ref{sec:overview}. For this we require the graph
    encoded by the table to be partially labeled. Concretely this means that the
    labels remain correct for boundary vertices when replacing one graph with a
    different one. We store all possible labels for boundary vertices, of which
    there are $b=O(\sqrt{\ell})$ many. In detail, when encoding a single graph $G$
    with $\ell$ vertices, we do not store a single unlabeled representative of $G$
    in the table (i.e., a graph isomorphic to all labeled versions of $G$), but
    all graphs such that $\ell-b$ vertices are unlabeled, e.g., have an arbitrary
    internal label, and $b$ vertices have all possible labelings in the range
    $0,\ldots, \ell$. This increases the size of the table by a negligible
    factor, outined now. Due to the partial labeling we require,
    the number of bits needed to store an index
    into the table increases by $O(\log({\ell \choose b}))=O(\log({\ell \choose \sqrt{\ell}}))=o(\ell)$,
    and thus is negligible for our use-case. Later, when we modify a graph $G_i$
    encoded as an index $i$ of lookup table (e.g., contract edges), we replace
    the index $i$ with the index $j$ such that the graph $G_j$ encoded by $j$
    represents the modified graph with the additional characteristic
    that all boundary vertices of $G_j$ have the same label in $G_i$. 
    the labeling for the non-boundary vertices changes due to
    this, but what we maintain is that a non-boundary vertex remains mapped to
    non-boundary vertex, and that all boundary vertices maintain their same labeling,
    which is all that we require for our data-structure. This is expressed
    via a set of invariants defined in Section~\ref{sec:dynmapping}
    and Section~\ref{sec:result}.
\lemtable

    For later alternative construction of our final encoding
    (Corollary~\ref{cor:consistentlabels}) and the final applications section
    (Section~\ref{sec:outer}) we require two variants of our lookup table.
    Firstly, we show that, if one wants to construct a table listing all labeled
    planar graphs with $\ell$ vertices, one requires $\mathcal{H}(\ell)+\ell \log \ell$ bits
    per index. 
    This follows from the simple calculation of the number of possible labeles
    of any graph with $\ell$ vertices.
    This is equal to the number of possible orderings of $[\ell]$, which is $\ell!$.
    And as $\log (\ell!) = O(\ell \log \ell)$, we arrive at the stated bit requirement
    per index. Recall that we later choose $\ell = \log^4 \log n$, so that the 
    the space usage for our use case remains small.
  \begin{corollary}
    \label{cor:labeledtable}
    The table of Lemma~\ref{lem:table} can be modified to store labeled planar
    graphs. Then each index into the table that references a graph with $\ell'$
    vertices requires $O(\ell' \log \ell')$ bits.
\end{corollary}

A final modification that we require in Section~\ref{sec:outer} is
storing colored graphs, such that vertices and/or edges are colored from a
constant-sized set of colors. For this we store all possible vertex- and edge 
colorings. For
each color this requires an additional $2^{O(\ell)}$ per index of table,
with $\ell$ the number of vertices of the respective graph encoded by the index.
This increases the bit requirement per index into the table by $O(\ell)$ bits
for a total space requirement of $\mathcal{H}(\ell)+o(\ell)+O(\ell)=O(\ell)$.

  \begin{corollary}
    \label{cor:coloredtable}
    The table of Lemma~\ref{lem:table} can be modified to store planar
    graphs that are vertex- and/or edge colored with a constant-sized set of colors.
    Then each index into the table that references a graph with $\ell'$
    vertices requires $O(\mathcal{H}(\ell'))=O(\ell)$ bits.
\end{corollary}
}

\sectable

\section{Succinct encoding of planar graphs}
\label{sec:blelloch}
We now describe the succinct encoding of unlabeled planar (and other
separable) graphs due to Blelloch and
Farzan~\cite{10.1007/978-3-642-13509-5_13}. We use their data structure as a
basis for our encoding. Our result effectively extend their encoding
with { (induced-) minor operations}. For this we need
to give a technical overview of their encoding. Let $G=(V, E)$ be an unlabeled
planar graph, $\mathcal{R}=\{P_1, \ldots, P_k\}$ an $r$-division with $r=\log^4
n$, and for each $P_i$ with $i \in [k]$, let $\mathcal{R}_i=\{P_{i, 1}, \ldots,
P_{i, k}\}$ be an $r'$-division of $P_i$ with $r'=\log^4 \log^4
n$.~\footnote{Blelloch and Farzan use $r'=\log n / \log \log n$ in their
publication, but make it clear that there is a large degree in freedom as long
as the choice is of size $o(\log n)$.} 
The encoding assigns three integer labels to each
vertex $u \in V$. A label in the entire graph (called \textit{global label}) a
label in each piece $P_{i} \in \mathcal{R}$ (called a \textit{mini label}) and a
label in each piece $P_{i, j} \in \mathcal{R}_i$ (called a \textit{micro
label}). We refer to $G$ with the newly assigned labels as the \textit{global
graph}, each labeled $P_i \in \mathcal{R}$ as a \textit{mini graph}, and each
labeled $P_{i, j} \in \mathcal{R}_i$ as a \textit{micro graph}. Note that
boundary vertices of $\delta \mathcal{R}$ receive multiple mini labels, and
analogous boundary vertices of $\delta \mathcal{R}_i$ receive multiple micro
labels. We refer to the boundary vertices of $\delta \mathcal{R}$ with their
assigned global labels as $\delta G$, and the set of boundary vertices of
$\delta \mathcal{R}_i$ with their assigned mini labels in $P_{i}$ as $\delta
P_{i}$. For a given boundary vertex $u$ identified by its global label, we refer
to all occurrences $u'$ (as a mini label) of $u$ in a mini graph $P_i$ as
\textit{duplicates}, and the same for boundary vertices of mini graphs $P_i$ in
regard to their occurrences in micro graphs. We refer to the set of (mini labels
of) duplicate vertices in a mini graph $P_{i}$ as $\Delta P_{i}$, and analogous
the set of (micro labels of) duplicate vertices in a micro graph $P_{i, j}$ as
$\Delta P_{i, j}$. 
Global labels are consecutive integers assigned first to all non-boundary vertices 
and then to boundary vertices, i.e., all boundary vertices have larger labels
than non-boundary vertices.
Analogous for mini labels in mini graphs. Micro labels are assigned arbitrarily.
For operations vertices are identified by their respective label.
E.g., a neighborhood query of a vertex $u \in V$ takes the global
label of $u$ as an input and outputs the global labels of all $v \in N(u)$,
analogous for queries in a mini or micro graph. Micro graphs
are encoded as an index into a lookup table $\mathcal{T}$, listing all
planar graphs of at most $r'$ vertices. Technically this is
realized by an array with one entry for each micro graph, which can be indexed
by $(i, j)$ when retrieving the entry for micro graph $P_{i, j}$. For our use
case we replace the table of Blelloch and Farzan with the table described in
Section~\ref{sec:table}, which provides additional operations. %It
%is easy to see that we can provide the label-preserving merge operation for
%every mini graph $P_{i, j}$ via our table.
We now describe operations that the encoding provides, which are
used by Blelloch and Farzan in their original publication, but are not defined
outside of 1. All mappings are implemented using IDs (Lemma~\ref{lem:ID})
over the universe $[n]$ combined with standard data structures such as lists
and pointers. Let $u \in V$ be a vertex identified by its global label. For
each such $u$, the encoding provides a mapping to access a list $\phi(u)$ that
contains tuples $(i, u')$ with $i$ the index of a mini graph $P_{i}$ that
contains mini label $u'$ of $u$. The lists are sorted in increasing order by
$i$. Note that if $u$ is a non-boundary vertex the mapping contains only a
single tuple. For each such tuple $(i, u')$ we can access a mapping
$\phi_i^{-1}(u')=u$. For vertices $u'$ in each mini graph $P_{i}$ (identified by
their mini label) analogous mappings $\phi_i(u')$ containing tuples $(j, u'')$
with $j$ the index of a micro graph $P_{i, j}$ that contains micro label $u''$
of $u'$, and the analogous mappings $\phi_{i, j}^{-1}(u'')=u'$ are provided. We
refer to all these mappings as \textit{static translation mappings}.

Recall that Blelloch and Farzan assign micro labels in an arbitrary fashion. We
instead assign the labels according to a coloring we define in the following.
Let $P_{i, j}$ be the micro graph we want to label. We first assign labels to
vertices that are neither a boundary vertex of $\delta \mathcal{R}$ nor of
$\delta \mathcal{R}_i$, which we assign the color \texttt{simple}. Then we
assign labels to vertices that are in the boundary $\delta \mathcal{R}$, but not
in $\delta \mathcal{R}_i$, which we assign the color \texttt{global-boundary}.
Then to vertices that are not in the boundary $\delta \mathcal{R}$, but in the
boundary $\delta \mathcal{R}_i$, colored \texttt{mini-boundary}, and finally to
vertices in both the boundary $\delta \mathcal{R}$ and in $\delta
\mathcal{R}_i$, colored \texttt{double-boundary}. Consequently, for any four
vertices of $P_{i, j}$ it holds that $a < b < c < d$ if $a$ is colored
\texttt{simple}, $b$ is colored \texttt{global-boundary}, $c$ is colored
\texttt{mini-boundary} and $d$ is colored \texttt{double-boundary}. For each
mini graph we store the lowest labeled vertex of each color, which uses
negligible space of $O((n / \log^4 \log^4 n) \log \log ^4 \log^4 n)=o(n)$ bits
overall.

Combined with our novel way of assigning micro labels to vertices of micro
graphs, the range-filtered neighborhood operation provided by our table allows
us to implement the \textit{color-filtered neighborhood} operation that outputs
all neighbors $x$ of a vertex $u$ {(in a micro graph $P_{i, j}$)} such that all
$x$ are colored with $c \in \{\texttt{simple}$, $\texttt{global-boundary},$
$\texttt{mini-boundary},$ $\texttt{double-boundary}\}$. The operation runs in
constant time (per element output).

Using the label preserving merge operation of the lookup table\full{\ from
Lemma~\ref{lem:table}} we can easily provide such merges for every micro graph.
Analogous for the vertex and edge deletion operation.

% Lastly, we provide what we call a \textit{duplicate re-label operation}. Given a
% duplicate vertex $u$, an index $i$ of a mini graph $P_{i}$ containing the
% micro label $u$ and any $x \in [n]$ the operation updates the mapping
% $\phi^{-1}_i(u):=x$. Analogous for duplicate vertices in a micro graph. As this
% mapping is realized as an array, realizing this is trivial.
For a given unlabeled planar graph $G$ we refer to the encoding described in
this section as \textit{basic encoding of $G$}. Kammer and
Meintrup~\cite{cloud22} have shown that the encoding can be constructed in
$O(n)$ time using $O(n)$ bits\full{\ during the construction phase when an
unlabeled simple planar graph is given as the input}.
Our modifications have negligible impact on the runtime and space usage of the
construction. This results in the following theorem.
% A short proof regarding the
% construction can be found in Appendix~\ref{sec:technical}.

\begin{theorem}
    \label{thm:basicenc}
    Let $G$ be an unlabeled planar graph and $\mathcal{H}(n)$ the entropy of
    encoding a planar graph with $n$ vertices. There exists a basic encoding of
    $G$ into a global graph, mini graphs and micro graphs that uses
    $\mathcal{H}(n)+o(n)$ bits total. The basic encoding provides static
    translation mappings for the global graph, each mini graph and each
    micro graph. 
    % Moreover, it provides duplicate re-label operations for any
    % duplicate vertex in mini  and micro graphs. 
    For each micro graph the encoding provides degree, adjacency,
    (color-filtered) neighborhood, (batch) edge/vertex deletion and
    label-preserving merge operations in $O(1)$ time. The basic encoding can be
    constructed in $O(n)$ time using $O(n)$ bits.
\end{theorem}

\section{Dynamic mapping data structures}
\label{sec:dynmapping}

For this section assume a planar graph $G$ is given via the basic encoding of
Theorem~\ref{thm:basicenc}. We now describe a set of dynamic mapping structures,
for which we outline the use-case in the following. We already mentioned in
Section~\ref{sec:overview} that a vertex that is initially a boundary vertex
(globally or/and in mini graphs) will never become a non-boundary vertex, and a
non-boundary vertex will never become a boundary vertex due to any of our edge
contractions. We construct dynamic variants of the static translation mappings
for boundary vertices (in the global or in mini graphs). We ensure that the
static translation mappings remain valid for all non-boundary vertices. Later
these mappings are concretely constructed for the vertices of the initial graph
(i.e., before any contractions are processed) and are maintained for all of
these vertices throughout. Concretely this means that the sets for which we define 
mappings and data structures never change after initialization. Recall
from Section~\ref{sec:overview} that when contracting an edge $\{u, v\} \in E$
we effectively forward the contraction operation to mini graphs that contain
mini labels $u'$ and $v'$ of $u$ and $v$ respectively, and then forward
the contraction to micro graphs in an analogous way. For our solution we require
that these cascading merge operations are handled independently without
interfering with each other.

Consider for example the case where we contract an edge $\{u, v\} \in E$ with
$u$ being a boundary vertex and $v$ a non-boundary vertex. In this case we want
to contract $v$ to $u$ (technical reasons for this are outlined in the next
section). To fulfill this contraction, we forward a request to the mini graph
$P_i$ to merge the vertices $u'$ and $v'$, the respective mini labels of $u$ and
$v$ in $P_i$. In the case that $v'$ is a boundary vertex in $P_i$, but $u'$ is a
non-boundary vertex in $P_i$, we want to merge $u'$ to $v'$, which is in
conflict with our desire to merge $v$ to $u$ in the global graph. The idea is to
support free-assignment merges, whose realization is described in the next
paragraph. This sort of conflict only pertains to vertices $u$ that are part of
the boundary $\delta G$ (thus, a duplicate $\Delta P_i$ in $P_i$) and/or have a
mini label $u'$ in some $P_i$ that is part of the boundary $\delta P_i$, i.e.,
we need to provide free-assignment merges for vertices of $\delta P_i \cup
\Delta P_i$. We construct a dynamic mapping that allows us to
decide, when merging two vertices $u', v' \in \delta P_i \cup \Delta P_i$, if the vertex that
remains after the merge is labeled $u'$ or $v'$. This is realized by assigning
each such vertex an \textit{external mini label} and an \textit{internal mini
label}. Effectively we have no free choice on which internal mini label
the vertex has after a merge, but we are free to assign a new external label. We
construct a mapping $\texttt{internal}_i: \delta P_i \cup \Delta P_i \rightarrow
\delta P_i \cup \Delta P_i$ that maps a given external label of a vertex of
$\delta P_i \cup \Delta P_i$ to its internal label, and a mapping
$\texttt{external}_i: \delta P_i \cup \Delta P_i \rightarrow \delta P_i \cup
\Delta P_i$ that maps a given internal label of a vertex $\delta P_i \cup
\Delta P_i$ to its external label. For all other vertices of $P_i$ we ensure
that the external and internal mini labels are identical, and therefore do not
need to construct any mapping. This is explicitly defined in
Invariant~\ref{inv:1} later in this section.

\newcounter{countlemmappingone}
\newcommand{\lemmappingone}{
  \setcounter{countlemmappingone}{\thetheorem}
  \begin{lemma}
    \label{lem:mappingA}
    All mappings $\texttt{internal}_i$ and $\texttt{external}_i$ can be constructed
    in $O(n)$ time using $o(n)$ bits of space. They provide read/write access in
    $O(1)$ time.
  \end{lemma}
}

\newcommand{\proofmappingone}{
\begin{proof}
    The mapping $\texttt{internal}$ is realized as follows. For each mini graph
    $P_i$ construct an ID (Lemma~\ref{lem:ID}) $I_i$ over the universe $V(P_i)$
    which manages the set $\delta P_i \cup \Delta P_i$. Additionally, construct
    an array $A_i$ that stores for each $u \in \delta P_i \cup \Delta P_i$ at
    index $I_i.\texttt{rank}(u')$ a value from universe $\delta P_i \cup \Delta
    P_i$. Each ID is constructed in linear time and the bits required for all
    IDs is $o(n)$. The arrays use $o(n)$ bits in total as they store $O(n/\log^2
    \log^4 n)$ entries, each of size $O(\log \log^4 n)$ bits and are initialized
    in $O(n)$ time. To read a value $\texttt{internal}_i(u')$ for $u' \in \delta
    P_i \cup \Delta P_i$ simply read the value stored at $A_i[x]$ with
    $x=I_i.\texttt{rank}(u')$ and to write the value
    $\texttt{internal}_i(u'):=v'$ for some $v' \in \delta P_i \cup \Delta P_i$
    simply set $A_i[x]:=v'$. Clearly this runs in constant time. For the inverse
    mappings $\texttt{external}_i$ we construct the same data structures with
    the same runtime and bits.
\end{proof}
}
\lemmappingone
\full{\proofmappingone}

We also have the need for a dynamic version of the static translation mappings
$\phi$ and $\phi_i$ for boundary vertices $\delta G$ in the global graph and
boundary vertices $\delta P_i$ in mini graphs, and the mappings $\phi^{-1}_i$
and $\phi^{-1}_{i, j}$ for the duplicate vertices $\Delta P_i$ in mini graphs
and $\Delta P_{i, j}$ micro graphs, outlined in the following paragraphs.
Initially these are equal to the static mappings. We first describe the dynamic
versions of the mappings $\phi^{-1}_i$ and $\phi^{-1}_{i, j}$, which we refer to
as $\Phi^{-1}_i$ and $\Phi^{-1}_{i, j}$ respectively. Afterwards we describe the
dynamic versions of the mappings $\phi$ and $\phi_{i}$, referred to as $\Phi$
and $\Phi_{i}$. To give an intuition for the use-case of these mappings,
consider a contraction of an edge $e=\{u, v\} \in E$ with $u, v \in \delta G$.
We contract this edge by first merging all $u'$ and $v'$ in the mini graphs
$P_i$ that contain both the duplicate $u'$ of $u$ and $v'$ of $v$. In all $P_i$
that contain only a duplicate of $v'$ of $v$ we need to know that the global
label of $v'$ is now (i.e., after the contraction) $u$ instead of $v$, for which
we use the described mappings. I.e., we insert the tuple $(i, v')$ to $\Phi(u)$
and set $\Phi^{-1}_i(v'):=u$. In all other mini graphs no changes need to be made.
\newcounter{countlemmappingtwo}%
\newcommand{\lemmappingtwo}{%
  \setcounter{countlemmappingtwo}{\thetheorem}%
  \begin{lemma}
    \label{lem:mappingB}
    All mappings $\Phi^{-1}_i: \Delta P_i \rightarrow \delta G$ and $\Phi_{i,
    j}^{-1}: \Delta P_{i, j} \rightarrow \delta P_i$ can be constructed in
    $O(n)$ time using $o(n)$ bits of space. They provide read/write access in
    $O(1)$ time.
  \end{lemma}
}
\newcommand{\proofmappingtwo}{
\begin{proof}
    The mappings are realized as follows. For each mini graph $P_i$ construct an
    ID $I_i$ (Lemma~\ref{lem:ID}) over the universe $V(P_i)$ which manages the
    set $\Delta P_i$. Additionally, construct an array $A_i$ that stores for
    each $u \in \Delta P_i$ at index $I_i.\texttt{rank}(u)$ any value from
    universe $\delta G$. Each ID is constructed in linear time and the bits
    required for all IDs is $o(n)$. The IDs and arrays use $o(n)$ bits in total
    as they store $O(n/\log^2 n)$ entries, each of size $O(\log n)$ bits and are
    initialized in $O(n)$ time. To read a value $\Phi^{-1}(u')$ for $u' \in
    \Delta P_i$ simply read the value stored at $A_i[x]$ with
    $x=I_i.\texttt{rank}(u')$ and to write the value $\Phi^{-1}(u'):=v$ for some
    $v \in \Delta G$ simply set $A_i[x]:=v$. Clearly this runs in constant time.
    For the analogous mappings $\Phi^{-1}_{i, j}$ constructed for all micro
    graphs $P_{i, j}$ construct the same data structures. All IDs and arrays
    constructed for the micro graphs store $O(n/\log^2 \log^4 n)$ entries, each
    of size $O(\log \log^4 n)$, i.e., $o(n)$ bits total.
\end{proof}
}
\lemmappingtwo
\full{\proofmappingtwo}

We now describe the dynamic mappings $\Phi$ and $\Phi_i$. As mentioned, we
initially require all mappings $\Phi(u)$ to be equal to $\phi(u)$ for $u \in
\delta G$ and analogously all mappings $\Phi_i(u')$ to be initially equal to
$\phi_i(u')$ for $u' \in \delta P_i$ for all mini graphs $P_i$. To represent
these mappings we construct a graph $H$ that contains all boundary vertices $u
\in \delta G$ and, for each mini graph $P_i$, a vertex $p_i$, with edges $\{u,
p_i\}$ added to $H$ exactly if $u$ has a duplicate $u'$ in $P_i$. 
Note that the existance of a duplicate $u'$ in $P_i$ means that $u'$
has a non-boundary neighbor in $P_i$ (initially). At each such
edge we store the tuple $(i, u')$. {We construct $H$ using the
forbidden vertex graph data structure of Lemma~\ref{lem:bvm}.} The tuples stored
at the incident edges of a vertex $u \in \delta G$ in $H$ are exactly the set
$\phi(u)$. Note that $H$ is a minor of $G$ and therefore planar. We
can provide for all $u \in \delta G$: iteration over all elements $\Phi(u)$ (by
iterating over $N(u)$ in $H$), insertion and removal of elements in $\Phi_i(u)$
(by inserting or removing edges in $H$), the merge of two sets $\Phi(u)$ and
$\Phi(v)$ for some $v \in \delta G$ (by merging $u$ to $v$ or $v$ to $u$ in
$H$). Some other similar operations are provided, outlined in detail in the next
section where we concretely describe our edge contraction algorithm. For each
mini graph $P_i$ the analogous graph $H_i$ is constructed, which manages the
mappings $\Phi_i$. An important note is that for each tuple $(i, u') \in
\Phi(u)$ for $u \in \delta G$ the vertex $u'$ is the external mini label of some
vertex in $P_i$. As no external or internal micro labels are defined for micro
graphs, each tuple $(j, u'') \in \Phi_i(u')$ for all mini graphs $P_i$ contains
the concrete micro label $u''$ in $P_{i, j}$. In the next section we describe
our neighborhood operation, for which we require a special version of the
mappings $\Phi$, which we first motivate with an intuition. To output the
neighborhood of a vertex $u \in \delta G$ we (intuitively) iterate over all $(i,
u') \in \Phi(u)$ and, for each such $(i, u')$, iterate (and translate to global
labels) over all neighbors of $u'$ in $P_{i}$. To achieve a runtime of
$O(|N(u)|)$ for this operation, we require that each tuple $(i, u')$
"contributes" at least one such neighbor. While this is true initially, due to
edge contractions {(and other modifications)} the degree of each such
$u'$ can become $0$. To remedy this, we store a special version of the mappings
$\Phi(u)$ which we refer to as $\Phi^{>0}$, containing only tuples $(i, u') \in
\Phi(u)$ such that the degree of $u'$ is $>0$. We construct the analogous
mappings $\Phi^{>0}_i$ for all $P_i$. During contractions, we update the mappings
$\Phi^{>0}$ and $\Phi^{>0}_i$ to uphold the aforementioned characteristic, which
is formalized in Invariant~\ref{inv:nonzero}. How this invariant is upheld, is
discussed in the next section. We realize these mappings exactly as $\Phi$ and
$\Phi_i$, respectively, i.e., as graphs $H^{>0}$ and $H^{>0}_i$. 
Initially $H^{>0}=H$ and all $H_i^{>0}=H_i$, by the definition of boundary
vertices in $r$-divisions (Section~\ref{sec:pre}). 

\begin{invariant}[non-zero-degree invariant]
    \label{inv:nonzero}
%    \textbf{(Non-zero-degree invariant)}
    For all $u \in \delta G$, each entry $(i, u') \in \Phi^{>0}(u)$ guarantees
    that $u'$ has degree $>0$ in mini graph $P_i$. For all $u' \in \delta P_i$
    over all mini graphs $P_i$, each entry $(j, u'') \in \Phi^{>0}_i(u')$
    guarantees that $u''$ has degree $>0$ in micro graph $P_{i, j}$.
\end{invariant}
\newcounter{countlemmappingthree}%
\newcommand{\lemmappingthree}{%
  \setcounter{countlemmappingthree}{\thetheorem}%
\begin{lemma}
    \label{lem:mappingC}
    Graphs $H$, $H^{>0}$ and $H_i$, $H^{>0}_i$ can be
    constructed in $O(n)$ time and $o(n)$ bits.
\end{lemma}}
\newcommand{\proofmappingthree}{
\begin{proof}
    The graph $H$ contains $O(n /\log^2 n)$ nodes as per construction. The data
    structure of Lemma~\ref{lem:bvm} is used to construct and manage $H$, which
    results in a construction time of $O(n/\log n)=o(n)$ and the same number of
    bits. There are $O(n/\log^4 n)$ graphs $H_i$, each containing $O(\log^4 n /
    \log \log^4 n)$ vertices. This results in a total construction time of
    $O(n/\log^4n \cdot \log^4 n / \log^2 \log^4 n \cdot \log \log^4 n)=o(n)$ for
    all $H_i$ and the same number of bits. The graphs $H^{>0}$ and $H^{>0}_i$
    are constructed in the same way. To uphold Invariant~\ref{inv:nonzero}
    initially, simply do not add edges to $H^{>0}$ or $H^{>0}_i$ that would
    violate this invariant. 
\end{proof}
}
\lemmappingthree
\full{\proofmappingthree}

Using all data structures described in this section we uphold invariants below
while running contractions on $G$---the details on this are described in the
next section. For better readability we slightly abuse the definition of our
\texttt{internal} (\texttt{external}) mappings of by assuming they return the
identity function for $u' \notin \delta P_i \cup \Delta P_i$.

\begin{invariant}[label-translation invariants]
    \label{inv:1}
%    \textbf{(label-translation invariants)}
    \ 
    \begin{enumerate}[a.]
        \item \textbf{Global to external mini label and vice-versa}:
        \begin{enumerate}[I.]
            \item For each  $u \in V \setminus \delta G$ and $(i, u')=\phi(u)$ it
            holds that $u'$ is the external mini label of $u$ in $P_i$ and $\phi^{-1}(u')=u$.

            \item For each $u \in \delta G$ and $(i, u') \in \Phi(u)$ ($\Phi$ is
            the dynamic version of $\phi$) it holds that $u'$ is the external
            mini label of $u$ in $P_i$ and $\Phi^{-1}(u')=u$.
        \end{enumerate}

        \item \textbf{Internal to external mini label and vice-versa}:

        For each vertex $u' \in V(P_i)$ identified by its external mini label,
        it holds that $u^*=\texttt{internal}(u')$ is the internal mini label of
        $u'$ and $\texttt{external}(u^*)=u'$.

        \item \textbf{Mini to micro label and vice-versa}:
        \begin{enumerate}[I.]
            \item For each  $u' \in  V(P_i) \setminus \delta P_i$ (identified by its
            internal mini label) and $(j, u'')=\phi_i(u')$ it holds that $u''$ is
            the micro label of $u'$ in $P_{i, j}$ and $\phi_{i, j}^{-1}=u'$.

            \item For each $u^* \in \delta P_i$ (identified by its internal mini
            label) and for each $(j, u'') \in \Phi_i(u^*)$ it holds that $u''$ is
            the micro label of $u^*$ in $P_{i, j}$ and $\Phi_{i,
            j}^{-1}(u'')=u^*$.
        \end{enumerate}
    \end{enumerate}
\end{invariant}

\section{Towards a succinct dynamic encoding}
\label{sec:result}

%In this section we describe a data structure in conjunction with the succinct
%encoding described in the previous section that allows us to contract edges. 
For this section let $G=(V, E)$ be a graph encoded by the basic
encoding of Theorem~\ref{thm:basicenc}. Also, assume that the mappings
of Lemma~\ref{lem:mappingA},~\ref{lem:mappingB} and~\ref{lem:mappingC} are
constructed and available. In this section we describe our solution to support
modifications of $G$. We denote by $\bar{G}=(\bar{V}, \bar{E})$ the graph $G$
before any modifications are processed, e.g., contractions of edges. Analogously
we define by $\bar{P}_i$, $\bar{P}_{i, j}$ the initial mini and micro graphs,
respectively, with its initial vertices (as mini/micro labels). As sketched in
Section~\ref{sec:overview} we handle contractions between so-called boundary
edges with a \textit{boundary graph $F=G[\delta
G]$} and analogously a \textit{mini boundary graph $F_i=P_i[\delta P_i]$} for
each mini graph $P_i$. These graphs are realized via the forbidden vertex graph
data structure (Lemma~\ref{lem:bvm}). For $F$ we use as the set of forbidden
vertices the empty set. For each $F_i$ we use the duplicate vertices $\Delta
P_i$ of $P_i$ as the set of forbidden vertices.
During
initialization edges between forbidden vertices are removed. The
forbidden-vertex graph data structure makes sure that this remains true after
initialization. Let $\{u, v\} \in E$ be an edge. We say $\{u, v\}$ is
\textit{managed by $F$} if $\{u, v\} \in E(F)$, $\{u, v\}$ is \textit{managed by
$F_i$} if $\{u', v'\} \in E(F_i)$ with $u'$ and $v'$ the mini labels of $u$ and
$v$ respectively, and finally we say $\{u, v\}$ is \textit{managed by $P_{i,
j}$} if it contains the edge $\{u'', v''\}$ with $u'', v''$ being the micro
labels of $u$ and $v$, respectively. We uphold the following invariant: {
\begin{invariant}[edge-singleton invariant]
    \label{inv:2}
%    \textbf{(edge-singleton invariant)}
\
    \begin{enumerate}[a.]
        \item An edge $\{u, v\}$ is managed by $F$ exactly if $u, v \in
        \delta G$.
        \item An edge $\{u, v\}$ is managed by $F_i$ exactly if $u',
        v' \in \delta P_i \setminus \Delta P_i$, with $u', v'$ the respective
        mini labels of $u$ and $v$ in $P_i$. In this case, no other $F_j$
        ($j\neq i$) also manages $\{u, v\}$.
        \item All edges $\{u, v\}$ not managed by $F$ or some $F_i$ are managed
        by one micro graph $P_{i, j}$.
    \end{enumerate}
\end{invariant}}\ 

Our construction of $F$ and each $F_i$ is the first step to achieve this
invariant. We now give an intuition why this invariant is useful. Due to some
edge contractions new edges $\{u, v\}$ might occur in $G$ between boundary
vertices (either global boundary vertices or vertices that are boundary vertices
in a mini graph). We can not afford to add this edge to all mini and micro
graphs that contain mini and micro labels of both $u$ and $v$, respectively.
Instead, we only add this edge to $F$ or some $F_i$. Moreover, if edges $e
\in E$ are managed multiple times, the runtime of the neighborhood
operation can increase.

An important note is that the basic encoding of $G$ does not adhere to the
edge-singleton invariant from the get-go, i.e., edges managed by some $F$ or
$F_i$ might be contained in one or more micro graphs initially. Using the batch
edge deletion operation provided for micro graphs (Theorem~\ref{thm:basicenc})
we can delete all edges that would initially violate our invariant. If this violates Invariant~\ref{inv:nonzero}, we
remove the respective entries from $\Phi^{>0}$ ($\Phi^{>0}_i$).
This uses $O(n)$ time.

We refer to the combination of the basic encoding of $G$, the mappings of
Lemma~\ref{lem:mappingA}, Lemma~\ref{lem:mappingB} and Lemma~\ref{lem:mappingC},
the boundary graph $F$ and each mini boundary graph $F_i$ as \textit{succinct
dynamic encoding of $G$}, summarized in the following corollary.

\begin{corollary}
    \label{lem:initialization}
    The succinct dynamic encoding of $G$ can be constructed in $O(n)$ time using
    $O(n)$ bits. After construction the encoding requires $\mathcal{H}(n)+o(n)$
    bits. The encoding upholds the label-translation, edge-singleton and
    non-zero degree invariant.
\end{corollary}

We now give an intuition how we implement the neighborhood operation for a
vertex $u \in V$ identified by its global label. We first output all neighbors
of $u$ in $F$ (which is $\emptyset$ if $u$ is not a boundary vertex) and then,
for all $P_i$ that contain a mini label $u'$ of $u$, compute all neighbors $v'$
of $u'$ in $F_i$ (which is again $\emptyset$ if $u'$ is not a boundary vertex) and
output the respective global label $v$ of $v'$. We then go to all micro graphs
$P_{i, j}$ that contain a mini label $u''$ of $u'$, compute all neighbors $v''$
of $u''$ in $P_{i, j}$ and output the respective global label $v$ of $v''$. By
the edge singleton invariant it is easy to see that each neighbor $v$ of $u$ in
$G$ is output exactly once by this algorithm. Invariant~\ref{inv:1} provides
the necessary translation operations. The missing details are
discussed in the proof of the following lemma.

\newcounter{countlemneighborhood}
\newcommand{\lemneighborhood}{
  \setcounter{countlemneighborhood}{\thetheorem}
\begin{lemma}
    \label{lem:neighborhood}
    For any $u \in V$  the neighborhood operation runs in time $O(|N(u)|)$.
\end{lemma}}
\newcommand{\proofneighborhood}{
\begin{proof}
    We describe the neighborhood operation on three levels: neighborhood of a
    vertex in a micro graph, in a mini graph and in the global graph. First, to
    output all neighbors of a vertex $u''$ in a micro graph $P_{i, j}$ simply
    use the neighborhood operation provided for micro graphs
    (Theorem~\ref{thm:basicenc}). This takes constant time per element output.
    To output the neighborhood of a vertex $u^* \in V(P_i)$, identified by its
    internal mini label, we distinguish between two cases: {(NM1)}
    $u^* \in \delta P_i$ and {(NM2)} $u^* \notin \delta P_i$. For
    Case NM1 proceed as follows: first output all neighbors of $u^*$ in $F_i$.
    Then, for each $(j, u'') \in \Phi^{>0}_i(u^*)$ iterate over all neighbors
    $v''$ of $u''$ in $P_{i, j}$. Translate each such $v''$ to its respective
    internal mini label $v^*$ with either $\Phi_{i, j}^{-1}(v'')=v^*$ if $v''
    \in \Delta P_{i, j}$, or with $\phi_{i, j}^{-1}(v'')=v^*$ if $v'' \notin
    \Delta P_{i, j}$, as per Invariant~\ref{inv:1}. Due to
    Invariant~\ref{inv:nonzero}, iterating over all $(j, u'') \in
    \Phi^{>0}_i(u^*)$ runs in $O(d_{u^*})$ time with $d_{u^*}$ the degree of
    $u^*$ in $P_{i, j}$. Translating and outputting all neighbors of all $u''$
    in all $P_{i, j}$ can be done in constant time per element output.
    Outputting all neighbors of $u^*$ in $F_i$ runs in constant time per element
    output as well. In Case NM2 simply determine the micro graph $P_{i, j}$ that
    contains the micro label $u''$ of $u^*$ via $\phi_i(u')=(j, u'')$
    (Invariant~\ref{inv:1}). Translating and outputting all neighbors of $u''$
    in $P_{i, j}$ is analogous to Case NM1. The runtime is again constant per
    element output. This is sufficient to provide a neighborhood operation for
    any $u^* \in V(P_{i})$ in constant time per element output. To provide the
    neighborhood operation for a vertex $u \in V$ we again distinguish between
    two cases: (NG1) $u \in \delta G$ and (NG2) $u \notin \delta G$. For Case
    NG1 we first output all neighbors of $u$ in $F$. Secondly, for each $(i, u')
    \in \Phi^{>0}(u)$ we determine the internal label $u^*$ of $u'$ in $P_i$ via
    $\texttt{internal}[u']=u^*$, iterate over all neighbors $v^*$ of $u^*$ in
    $P_i$ and translate each $v'=\texttt{external}[v^*]$ to its global label $v$
    via $\Phi_i^{-1}(v')=v$ (if $v' \in \Delta P_i$) or $\phi_i^{-1}(v')=v$ (if
    $v' \notin \Delta P_i$). All reads of the various mappings are correct as
    per Invariant~\ref{inv:1}. The runtime analysis works analogous to Case NM1
    in mini graphs, i.e., constant time per element output. For Case NG2 simply
    determine the mini graph $P_i$ that contains a duplicate $u'$ of $u$ via
    $\phi(u)=(i, u')$ and continue analogous to Case NG1. The correctness
    follows again from Invariant~\ref{inv:1}. The runtime is constant per
    element output. Due to Invariant~\ref{inv:4} we have output each neighbor of
    $u$ exactly once.
    \end{proof}
}
\lemneighborhood
\full{\proofneighborhood}

We now focus on our edge-contraction algorithm. For this we introduce one last
invariant, which we call the \textit{status invariant}. As sketched in
Section~\ref{sec:overview} we require that for every vertex being a boundary
vertex (either globally or in a mini graph) to remain a boundary vertex, and for
every non-boundary vertex to remain a non-boundary vertex. For this we slightly
abuse the definition of boundary vertices. By definition of $r$-divisions
(Section~\ref{sec:pre}) a boundary vertex $u \in \delta G$ has
neighbors in more than one mini graph. Due to contractions {(or other
modifications)} this might at some point no longer be true. Nonetheless, we
still consider such a vertex to be a boundary vertex. We require that a boundary
vertex remains a boundary vertex, and a non-boundary vertex remains a
non-boundary vertex. For this we maintain the following invariant that depends
on our slight abuse of the boundary vertex definition.

\begin{invariant}[status invariant]
    \label{inv:4}
    For every $u\in V$ and every $u' \in V(P_i)$ over all mini graphs $P_i$, it
    holds $u \in \delta G$ if and only if $u \in \delta \bar{G}$ as well as $u'
    \in \delta P_i$ if and only if $u' \in \delta \bar{P}_i$.
\end{invariant}

We now give an overview of our edge-contraction algorithm, which we describe in
three levels: vertex merges in micro graphs, vertex merges in mini graphs and
edge contractions in the global graph. We guarantee the four invariants
(Invariants~\ref{inv:nonzero},~\ref{inv:1},~\ref{inv:2} and~\ref{inv:4}) before
and after each edge contraction. Technically, merges are executed in (mini)
boundary graph(s) and micro graphs. Everything else is to maintain the mappings
of Section~\ref{sec:dynmapping}. An edge $\{u, v\} \in E$ is contracted by
determining all micro graphs $P_{i, j}$ that contain (micro labels of) $u$ and
$v$, all mini boundary graphs $F$ that contain (mini labels of) $u$ and $v$ and
check if $F$ contains $u$ and $v$. In all structures that contain $u$
and $v$ we merge $v$ to $u$ and update the mappings of
Section~\ref{sec:dynmapping}.\full{\ An important fact we make use of is
that during this contraction algorithm anytime a pair of vertices is merged in a
mini or micro graph, an edge between this pair exists that is managed exactly
once (Invariant~\ref{inv:2}).} To guarantee the invariants we split the
responsibilities among the three levels:

\begin{itemize}
    \item \textbf{Global Graph-Responsibility.} Contractions in the global
    graph maintain Invariant~\ref{inv:nonzero} regarding
    $\Phi^{>0}$, Invariant~\ref{inv:1}.a., {Invariant~\ref{inv:2}.a
    and Invariant~\ref{inv:4} for all $u \in V$.

    \item \textbf{Mini Graph-Responsibility.} Vertex merges in a mini graph
    $P_i$ maintain Invariant~\ref{inv:nonzero} regarding
    $\Phi^{>0}_i$, Invariant~\ref{inv:1}.b-c, Invariant~\ref{inv:2}.b, and
    Invariant~\ref{inv:4} for all $u' \in V(P_i)$.

    \item \textbf{Micro Graph-Responsibility.} Vertex merges in a micro graph
    maintain Invariant~\ref{inv:2}.c.}
\end{itemize}

Our contraction algorithm is built up from bottom-to-top, i.e., we first
describe merges in micro graphs, then mini graphs (and mini boundary graphs) and
edge contractions in $G$ (and merges in $F$). To uphold the
responsibilities of micro graphs we implement a variant of the
forbidden-vertex graph data structure (Lemma~\ref{lem:bvm}) for micro graphs,
summed up in the following lemma. To uphold Invariant~\ref{inv:4} we
are not allowed to merge a vertex $v''$ to a vertex $u''$ in a micro graph
$P_{i, j}$ if $v'' \in \Delta P_{i, j}$ and $u'' \notin \Delta P_{i, j}$, which
we formulate explicitly.

\newcounter{countlemmicromerge}
\newcommand{\lemmicromerge}{
  \setcounter{countlemmicromerge}{\thetheorem}
\begin{lemma}
    \label{lem:micromerge}
    For all micro graphs $P_{i, j}$ we can provide free
    assignment merges for each $P_{i, j}$ such that no edges $\{u'', v''\}$
    exists that should be managed by $F$ or $F_i$. If such an edge would occur
    due to the merge, it is not inserted to $P_{i, j}$ and instead returned.
    Computing any number of such merges among all micro graphs can be done in
    $O(n)$ total time. The operation upholds the micro graph-responsibility.
    Merging a vertex $v''$ to a vertex $u''$ is {not allowed} if $v''
    \notin \Delta P_{i, j}$ and $u'' \in \Delta P_{i, j}$. All other (planar
    preserving) merges are allowed.
\end{lemma}
}
\newcommand{\proofmicromerge}{
\begin{proof}
    In this proof we discuss how to uphold Invariant~\ref{inv:2}.c regarding
    micro graphs, i.e., the micro graph responsibility. Denote with $u'', v''$
    the two vertices we want to merge in micro graph $P_{i, j}$, with $u'$ and
    $v'$ the mini labels of $u''$ and $v''$ in $P_{i}$, respectively.\ W.l.o.g.,
    assume we want to merge $v''$ to $u''$. Recall from
    Section~\ref{sec:blelloch} that the vertices in micro graphs are assigned
    colors $\mathcal{C}=\{\texttt{simple}$, $\texttt{global-boundary},$
    $\texttt{mini-boundary},$ $\texttt{double-boundary}\}$ and that we are able
    to iterate over all neighbors of a given color in constant time per element output
    via the color-filtered neighborhood operation. As a refresher, we outline
    the meaning of the coloring again. Let $u''$ be a vertex in $P_{i, j}$.
    Denote with $u'$ the mini label of $u''$ in $P_i$ and with $u$ the global
    label of $u'$ in $G$. The vertex $u''$ is colored \texttt{simple} exactly if
    $u' \notin \delta P_i$ and $u \notin \delta G$. It is colored
    $\texttt{global-boundary}$ exactly if $u' \notin \delta P_i$ and $u \in
    \delta G$. It is colored $\texttt{mini-boundary}$ exactly if $u' \in \delta
    P_i$ and $u \notin \delta G$. Finally, it is colored
    $\texttt{double-boundary}$ if it $u' \in \delta P_i$ and $u \in \delta G$.
    Due to Invariant~\ref{inv:2} we know that all edges that should be managed
    by $P_{i, j}$ must not be managed by $F_i$ or $F$. We denote a pair of
    vertices for which no edge is allowed to exist in $P_{i, j}$ as
    \textit{forbidden edges}. We can categorize forbidden edges $\{u'', v''\}$
    of $P_{i, j}$ by the coloring of $u''$ and $v''$. {The following
    combinations form a forbidden edge: (1) $u''$ is colored
    $\texttt{double-boundary}$ and $v''$ is colored with any $c' \in \mathcal{C}
    \setminus \{\texttt{simple}\}$ (or vice-versa) or (2) both $u''$ and $v''$
    are colored with the same color $c \in \mathcal{C} \setminus
    \{\texttt{simple}\}$. Note that this excludes pairs of vertices for which
    $u''$ is colored $\texttt{mini-boundary}$ and $v''$ is colored
    $\texttt{global-boundary}$ (or vice-versa), as clearly such an edge can not
    exist in either $F_i$ or $F$.} {We denote the set of all
    forbidden edges that exist in $P_{i, j}$ as actual edges with $E^{f}_{i,
    j}$.}

    To merge two vertices $u'', v'' \in V(P_{i, j})$ we first merge them using
    the merge operation provided for micro graphs due to
    Theorem~\ref{thm:basicenc}. Assume we merged $v''$ to $u''$. Note that
    before this merge it holds that $E^{f}_{i, j}=\emptyset$
    (Invariant~\ref{inv:2}) and after the merge it holds that $E^{f}_{i, j}=
    E^{f}_{i, j} \cap N(u'')$, i.e., all forbidden edges that now exist as
    actual edges are contained in the neighborhood of $u''$. Iterating over
    these forbidden edges in constant time per element can thus be reduced to a
    constant number of color-filtered neighborhood operations executed for
    $u''$. We use this to store $E^{f}_{i, j}$ in a list $L$ followed by
    removing $E^{f}_{i, j}$ from the graph using the batch edge-deletion
    operation. Afterwards we return the list $L$ as outlined in the lemma.

    We now discuss the runtime, for which we do a runtime analysis. Recall
    that we merge $v''$ to $u''$ by setting $N(u''):=(N(u'') \cup N(v''))
    \setminus \{\{v'', u''\}\}$. This means that we can view such a merge as
    changing one endpoint each of all edges incident to $v''$ to become incident
    to $u''$ instead. Self-loops and parallel edges are discarded during this
    process. Note that each edge of $P_{i, j}$ can become a forbidden edge
    only once, since we remove all such edges that occur from $P_{i, j}$.
    
    Due to the color-filtered neighborhood operation we only iterate over edges
    we insert to $L$. Thus the time to build all lists $L$ in one $P_{i, j}$ is
    bound by $O(|V(P_{i, j})|)$. Over all $P_{i, j}$ this sums up to $O(n)$ time
    among all $P_{i, j}$. 
    \end{proof}
}
\lemmicromerge
\full{\proofmicromerge}

We first note, whenever we call the merge operation of Lemma~\ref{lem:micromerge}
for a micro graph $P_{i, j}$ in the next paragraphs, the operation returns edges
$\{u'', v''\}$ that should be managed by $F_i$ or $F$, but not $P_{i, j}$. We
then translate $\{u'', v''\}$ to $\{u', v'\}$ with $u'$ and $v'$ the respective
mini labels of $u''$ and $v''$. If the edge $\{u', v'\}$ should be managed by
$F_i$, we insert it to $F_i$. Returned edges that should not be managed by $F_i$
are instead returned after the merge operation in $P_i$ is executed. This
upholds Invariant~\ref{inv:2} (restricted to micro and mini graphs). To uphold
Invariant~\ref{inv:nonzero} (for mini graphs) we check, after any call to a
merge of a vertex $v''$ to $u''$ in a micro graph $P_{i, j}$ if the degree of
$u''$ changed from $0$ to non-zero or vice-versa. If it does, we must possibly
update the mapping $\Phi^{>0}_i(u')$ to either include the tuple $(j, u'')$ or
remove it, with $u'$ the mini label of $u''$. Note that this is only done in the
case that $u'$ is a boundary vertex, as otherwise no mapping $\Phi^{>0}_i(u')$
exists. \full{The technical details regarding guaranteeing
Invariants~\ref{inv:nonzero} and~\ref{inv:2} is discussed in the proof of
Lemma~\ref{lem:minimerge}. For now, we
assume the necessary steps are executed during merges.}
% and for now assumed to be done by all merges.

Let $u', v' \in V(P_i)$ be two vertices identified by their external mini label.
To provide a merge of $u'$ and $v'$ we distinguish between three cases: (M1)
$u', v' \notin \delta P_i$, (M2) $u' \in \delta P_i$ and $v' \notin \delta P_i$
and (M3) $u', v' \in \delta P_i$. In Case M1 we determine the micro graph $P_{i,
j}$ that contains micro labels $u''$ and $v''$ of $u'$ and $v'$, respectively,
via the static mappings $\phi_i(u')=(j, u'')$ and $\phi_i(v')=(j, v'')$ as per
Invariant~\ref{inv:1}. In $P_{i, j}$ we merge $v''$ to $u''$ exactly if $v'$
should be merged to $u'$, and otherwise merge $u''$ to $v''$
(Lemma~\ref{lem:micromerge}). By this congruent choice of merge we uphold
Invariant~\ref{inv:1}.c without having to modify any mappings. Since the merged
vertex remains a non-boundary vertex, Invariant~\ref{inv:4} is guaranteed. This
concludes all responsibilities of merges in mini graphs. All operations take
constant time. Note that all merges of Case M1 are free-assignment merges.

Denote with $u^*=\texttt{internal}[u']$ and $v^*=\texttt{internal}[v']$ the
internal mini labels of $v'$ and $u'$ respectively. For Case M2 we are forced to
merge $v^*$ to $u^*$ to uphold Invariant~\ref{inv:4}, i.e., internally this
merge is not a free-assignment merge. To execute the merge we determine the
micro graph $P_{i, j}$ containing the micro label $v''$ of $v^*$ via
$\phi_i(v^*)=(j, v'')$. In $P_{i, j}$ we merge $v''$ to $u''$
(Lemma~\ref{lem:micromerge}) with $u''$ the micro label of $u^*$ in $P_{i, j}$.

Note that merging $u''$ to $v''$ is {not allowed}. We determine $u''$
via an operation we call \textit{micro-label search procedure}, which searches
for $u''$ by iterating over all $x'' \in N(v'') \cap \Delta P_{i, j}$ (the
neighbors of $v''$ that are duplicates) and testing if $\Phi^{-1}_{i,
j}(x'')=u^*$. If this is the case, we have found the duplicate $u'':=x''$ of
$u^*$. Note that this operation can fail, as $u''$ and $v''$ are not guaranteed
to be adjacent. In this {\textit{special case}}, we instead iterate over all
$x'' \in \Delta P_{i, j}$. A key characteristic to get a good runtime is that
{the special case only occurs} if the edges $\{u, v\}$ exists in $F$,
with $u$ and $v$ the global labels of $u'$ and $v'$, respectively, which allows
us to upper bound the number of {encountered} special cases by
$|E(F)|=O(n/\log^2n)$ times.

\newcounter{countlemsearchone}
\newcommand{\lemsearchone}{
  \setcounter{countlemsearchone}{\thetheorem}
\begin{lemma}
    \label{lem:search1}
    All executions of the micro-label search procedure are processed in $O(n)$ time over
    any number of vertex merges of Case M2 in all $P_i$.
\end{lemma}
}
\newcommand{\proofsearchone}{
\begin{proof}
    Let $u' \in \delta P_i$ and $v' \notin \delta P_i$ for some mini graph $P_i$
    and assume that we want to merge $v'$ to $u'$. As sketched, we determine the
    micro graph $P_{i, j}$ and the micro label $v''$ of $v'$ therein with $(j,
    v'')=\phi_i(v')$. Denote with $u''$ the micro label of $u'$ in $P_{i, j}$,
    which we are yet to determine. In the following if we say we \textit{check}
    if a vertex $x'' \in \Delta P_{i, j}$ is the micro label of $u'$ by testing
    if $\Phi_{i, j}^{-1}(x')$ equals $u'$, in which case we found the micro
    label $u''$ of $u'$. Clearly a single check takes $O(1)$ time. We analyze
    the runtime of the micro label search procedure in two cases: (1) $u'' \in
    N(v'')$ and (2) $u'' \notin N(v'')$. We always assume that Case 1 is true,
    and if it then turns out to be false, we fall back to Case 2. In Case 1 we
    procede as follows: {iterate over all elements $x'' \in N(v'')
    \cap \Delta P_{i, j}$ via the color-filtered neighborhood operation
    (Section~\ref{sec:blelloch}) and check if $\phi_{i, j}^{-1}(x'')=u'$, in
    which case we have found the micro label $u'':=x''$ of $u'$. If this search
    procedure was successful, we are done, otherwise continue to Case 2 and
    simply check all $x'' \in \Delta P_i$ in the same fashion, i.e., finding the
    $x''$ for which $\phi_{i,j}^{-1}(x'')=u'$ holds. We first analyze the
    runtime of Case 1. All $x'' \in N(v'') \cap \Delta P_{i, j}$ constitute
    edges $\{v'', x''\}$ that after the merge constitute edges $\{u'', x''\}$
    that are being discarded in $P_{i, j}$, and then managed by $F_{i}$. Thus,
    any edge in each $P_{i, j}$ participates in such a search procedure at most
    once and the total cost of searches of Case 1 is bounded by $O(n)$.}
    
    % For the
    % sake of analysis assume that we initially add a coin to each edge in all
    % micro graphs, i.e., directly after initialization of all structures. Merging
    % $v''$ to $u''$ can be thought of as re-directing all edges $\{v'', y''\}$ to
    % become edges $\{u'', y''\}$ with $y'' \in N(v'')$. Any edges that are
    % re-directed in a such a way due to any merge retain their coins, with
    % duplicate edges (or edges that should not be managed by $P_{i, j}$)
    % discarding their coins as they are being discarded. Assume anytime we
    % iterate over an $x'' \in N(v'') \cap \Delta P_{i, j}$ during the micro label
    % search procedure, we remove a coin from $\{x'', v''\}$. Once $v''$ is merged
    % to $u''$ all these edges are re-directed to be $\{x'', u''\}$, which are
    % then being discarded as they should not be managed by $P_{i, j}$. Therefore,
    % during any number of proceeding merges in $P_{i, j}$ there is never an
    % instance where a coin is taken from an edge without a coin. Thus, this
    % operation takes linear time in the number of edges over any number of merges
    % in all $P_{i, j}$, i.e., $O(n)$ total time. 
    
    As for Case 2, we only enter it exactly if we failed our search in Case 1.
    As we call the merge operation in a micro graph only as a sub-routine when
    contracting some edge $\{u, v\}$ in the global graph, this edge must be
    managed by either $F$, some $F_i$ or $P_{i, j}$ (Invariant~\ref{inv:2}). As
    we enter Case 2, we know it is not managed by $P_{i, j}$ (otherwise we would
    have found it during Case 1). {As we enter the micro-label
    search procedure only during case M2, i.e., when merging a non-boundary $v'$
    to a boundary $u'$ in $P_i$, we know the edge can not be managed by $F_i$.
    This means the edge must be managed by $F$ (as $\{u, v\}$).} We can
    therefore bound the number of times we enter Case 2 via the number of
    possible edge contractions executed on edges that are managed by $F$---which
    is $O(n/\log^2n)$. Any time we contract such an edge $\{u, v\} \in F$ we may
    iterate over all mini graphs $P_i$ that contain a duplicate $v'$ of $v$ and
    a duplicate $u'$ of $u$, where we then merge $v'$ to $u'$ (or vice-versa)
    for which the specified micro-label search procedure is executed as a
    subroutine in exactly one micro graph $P_{i, j}$. As the total number of
    duplicates among mini graphs is $O(n/\log^2 n)$, the number of such search
    procedures is $O(n/\log^2 n)$ as well. One run of such a search procedure
    can be done in $O(|V(P_{i, j})|)=O(\log^4 \log^4 n)$ time. This results in a
    total runtime of $o(n)$ over all micro label search procedures.
    \end{proof}
}
\full{\lemsearchone}
\full{\proofsearchone}

Once the merge is executed in $P_{i, j}$ we must possibly update the mappings
that translate between the internal and external mini labels. Recall that the
mappings $\texttt{internal}$ and $\texttt{external}$ are only available for
vertices of $\delta P_i \cup \Delta P_i$. In the case that $v' \in \delta P_i
\cup \Delta P_i$ we are able to provide a free assignment merge as follows: if
the request was to merge $u'$ to $v'$, we set $\texttt{internal}[v']=u^*$ and
$\texttt{external}[u^*]=v'$. Otherwise, no update is necessary. If $v' \notin
\delta P_i \cup \Delta P_i$ we are not able to provide a free assignment merge,
instead we are forced to merge $v'$ to $u'$. We refer to this situation as the
\textit{M2 special case}. If the merge was called with the request to merge $u'$
to $v'$, and we are in this M2 special case, the operation is not allowed. 
In our use case this case never arises.
Intuitively, constraints (e.g., Invariant~\ref{inv:4}) that force us to
contract $\{u, v\}$ by merging $v$ to $u$ {either "line up" with
being able (or forced) to merge of $v'$ to $u'$ in $P_i$, with $v'$ and $u'$
the mini labels of $v$ and $u$ in $P_i$, respectively, or if they do not line
up, we make use of the internal/external mappings.}

\full{As for Invariant~\ref{inv:1}.c, it is easy to see that it is still correct as we
have merged $v''$ to $u''$ and $v^*$ to $u^*$ so that we do not need to update
mappings $\Phi_i(u^*)$ and $\Phi_{i, j}^{-1}(u'')$. Our (possible) updates to
the internal/external mappings uphold Invariant~\ref{inv:1}.b, and our exclusion
of the invalid request for a free-assignment merge ensures that this can always
be done if needed. \full{Besides the micro-label search procedure, all
other operations take constant time per merge.}}

Finally, we consider Case M3. In this case both $u'$ and $v'$ are boundary
vertices with $u^*=\texttt{internal}[u']$ and $v^*=\texttt{internal}[v']$ being
the internal mini labels of $v'$ and $u'$ respectively. As sketched in
Section~\ref{sec:overview}, our intuition for merging $v'$ to $u'$ is that we
first merge all $v''$ to $u''$ in all micro graphs $P_{i, j}$ that contain both
a duplicate $u''$ of $u^*$ and $v''$ of $v^*$. Secondly, for all micro graphs
$P_{i, j}$ that contain only a duplicate $v''$ of $v^*$, but not of $u^*$, we
update the mappings $\Phi_{i, j}^{-1}(v''):=u^*$ and insert $(i, v'')$ to
$\Phi_i(u^*)$. Finally, we merge $v^*$ to $u^*$ in $F_i$. To describe the
realization technically we introduce some additional notation. Denote with
$Z_i^{u^* \cap v^*}$ the set of all triples $(j, u'', v'')$ with $(j, u'') \in
\Phi_i(u^*)$ and $(j, v'') \in \Phi_i(v^*)$, with $Z_i^{u^* \setminus v^*}$ the
set of all tuples $(j, u'')$ with $(j, u'') \in \Phi_i(u^*)$ such that no tuple
$(j, \cdot)$ is contained in $\Phi_i(v^*)$, and with $Z_i^{u^* \oplus v^*}$ all
tuples $(j, u'') \in \Phi_i(u^*)$ together with all tuples $(j', v'') \in
\Phi_i(v^*)$ for which it holds that no tuple $(j', \cdot)$ exists in
$\Phi_i(u^*)$. To execute a merge of $v^*$ to $u^*$ in $P_i$, first iterate over
all triples $(j, u'', v'') \in Z_i^{u^* \cap v^*}$ and merge $v''$ to $u''$ in
$P_{i, j}$, then iterate over all tuples $(j, v'') \in Z_i^{v^* \setminus u^*}$
and update all mappings $\Phi_{i, j}^{-1}(v''):=u^*$. Finally, set
$\Phi_i(u^*):=Z_i^{u^* \oplus v^*}$ and merge $v^*$ to $u^*$ in $F_i$.
\conf{In the proof we show how these sets occur
(intuitively) "naturally" via merges in the graph $H_i$, which manages $\Phi_i$.}

\newcommand{\extendedmthirdcase}{
As mentioned in Section~\ref{sec:dynmapping}, each $\Phi_i$ is implemented
as a graph $H_i$ using the forbidden vertex data structure (Lemma~\ref{lem:bvm})
such that for all $u \in \delta P_i$, the auxiliary data stored at edges
incident to $u$ in $H_i$ make up the set $\Phi_i(u)$, with each edge
corresponding to exactly one entry in $\Phi_i(u)$. Computing the previously
defined sets and updating the mappings is done automatically when merging $v^*$
to $u^*$ in $H_i$: Setting $\Phi_i(u^*):=Z_i^{u^* \oplus v^*}$ is exactly the
merge of $v^*$ to $u^*$ in $H_i$. The sets $Z_i^{v^* \setminus u^*}$ are exactly
the edges that are inserted to $N(u^*)$ in $H_i$ due to the merge, and the set
$Z_i^{u^* \cap v^*}$ is exactly the set of parallel edges that are being
discarded. All these sets are output due to a merge of $u^*$ and $v^*$ in
$H_i$, as outlined in Section~\ref{sec:pre}. Even though both $H_i$ and $F_i$
provide free assignment merges, we are forced to merge $v^*$ to $u^*$ exactly if
$|\Phi_i(u^*)| > |\Phi_i(v^*)|$, as otherwise we are not able to achieve our
runtime goal of $O(n)$ time, with the bottleneck being the computation and
iteration over the set $Z_i^{v^* \setminus u^*}$. Since we have the mappings
$\texttt{internal}/\texttt{external}$ and merges in $F_i$ are always free
assignment, this causes no problems for us. We simply update the mappings
$\texttt{internal}$ and $\texttt{external}$ for $u^*$, $v^*$ and $u', v'$
analogous to Case M2, ensuring that the merge of $u'$ and $v'$ is a free
assignment merge as well. Thus, Invariant~\ref{inv:1}.b holds.
Invariant~\ref{inv:1}.c is managed by the update of $\Phi_i(u^*)$, which
correctly inserts all needed values due to merging of $v^*$ to $u^*$ in $H_i$.
Regarding Invariant~\ref{inv:4}, as we merge two boundary vertices the vertex
that remains after the merge is still a boundary vertex. Denote with
$n_i$ the number of vertices in $H_i$. Due to the use of Lemma~\ref{lem:bvm} for
$H_i$, any number of merges in $H_i$ can be processed in $O(n_i \log^2 n_i)$
time. Computing and iterating over all elements of the sets $Z_i^{u^* \cap
v^*}$, $Z_i^{v^* \setminus u^*}$, and computing the sets $Z_i^{u^* \oplus v^*}$
due to any number of merges of Case M3 can be done in $O(n_i \log^2 n_i)$ time
as well. For each element in the computed sets, a constant number of operations
are executed resulting in $O(|E(H_i)|)=O(n_i)$ time. Over all $H_i$ this sums up
to a total runtime of $O(n)$. What we have yet to address is updating the
mapping $\Phi^{>0}_i(u^*)$, which is done by simply merging $v^*$ to $u^*$ in
$H^{>0}_i$, with the same runtime as merging $v^*$ to $u^*$ in $H_i$.
}
\full{\extendedmthirdcase}
Combining Cases M1, M2 and M3 we show the following lemma.

\newcounter{countlemminimerge}
\newcommand{\lemminimerge}{
  \setcounter{countlemminimerge}{\thetheorem}
\begin{lemma}
    \label{lem:minimerge}
    All vertex merges in all mini graphs $P_i$ are processed in
    $O(n)$ time and uphold their responsibilities. Edges that would occur due to
    a merge in $P_i$, but should be managed by $F$ are returned. All merges
    excluding the M2 special case are free assignment merges.
\end{lemma}
}
\newcommand{\proofminimerge}{
\begin{proof}
    \conf{In this proof we first discuss the missing details regarding
    updating each $\Phi_i$ during merges of Case M3. Afterwards}\full{In this proof} we show how
     merges in a mini graph uphold Invariant~\ref{inv:nonzero}
    and~\ref{inv:2}.
    
    \conf{\extendedmthirdcase}
    
    We \conf{now}\full{first} discuss upholding Invariant~\ref{inv:2}. Denote with $u', v' \in
    V(P_i)$ two vertices that should be merged, and assume that $v'$ is merged to
    $u'$. Denote with $u''$ and $v''$ the micro labels in a micro graph $P_{i, j}$
    of $u'$ and $v'$, respectively.
    
    Any time a merge of $u''$ and $v''$ in some $P_{i, j}$ is executed as a
    sub-routine, all edges $\{x'', y''\}$ that should not be managed by $P_{i,
    j}$ are returned, with $x'', y'' \in P_{i, j}$. Using the label translation
    mappings $\phi_{i, j}^{-1}$ or $\Phi_{i, j}^{-1}$ we translate $x'', y''$ to
    their respective mini labels $x'$ and $y'$. If $x', y' \in V(F_i)$ we first
    check if the edge $\{x', y'\} \in E(F_i)$. If this is not the case, we
    insert the edge. If one of $x'$ or $y'$ is not contained in $V(F_i)$ we know
    that the edge $\{x', y'\}$ should be managed by $F$. For this, we add all
    such $\{x', y'\}$ to a list $L$ which we return after the merge of $v'$ to
    $u'$ is executed. To analyze the runtime, note that in total there are at
    most $O(n)$ edges returned due to merges in micro graphs. This is due to the
    fact that all merges in micro graphs can be done in $O(n)$ time, which
    includes returning the aforementioned edges (Lemma~\ref{lem:micromerge}).
    For each such returned edge we first spend $O(1)$ time to translate its
    endpoints to their respective mini labels, and then $O(1)$ time to check if
    this edge is contained in $F_i$ (constant time adjacency queries are
    provided due to Lemma~\ref{lem:bvm}). We only call the insertion operation
    if this is not the case. The runtime of an insertion is logarithmic. Adding
    the edge to $L$ if it should not be managed by $F_i$ takes $O(1)$ time. As
    each $F_i$ is planar, at most a linear number of edges can be inserted to
    $F_i$ due to this, i.e., $O(\log^4n/\log^2 \log^4 n)$ edges can be inserted
    to each $F_{i}$. Each insert takes logarithmic time, resulting in a total
    runtime of $O(n/\log^4 n \cdot \log^4 n/\log^2 \log^4 n \cdot \log \log^4
    n)=O(n)$ over all $P_i$ and $F_i$ respectively.
    
    We now discuss how to uphold Invariant~\ref{inv:nonzero}. Let $u' \in V(F)$ and
    $(j, u'') \in \Phi^{>0}_i$. We first analyze the situations in which the degree
    of $u''$ switches between zero and non-zero.. First note that no merge of two
    vertices $a'', b'' \notin \Delta P_{i, j}$ can result in the degree of $u''$ to
    change (i.e., Case M1). Secondly, in the case we merge a vertex $v' \in V(F)$ to
    $u'$ (i.e., case M3), we know that the Invariant~\ref{inv:nonzero} is upheld
    without any extra operations as $u'$ is merged into $v'$ in $H_i^{>0}$ to update
    $\Phi^{>0}_i(u')$. After this merge, tuples $(j, u'')$ that are inserted to
    $\Phi^{>0}_i(u')$ indicate that $u'$ had no duplicate $u''$ with degree $>0$ in
    $P_{i, j}$ before the merge. Thus, we are only interested in Case M2, where we
    merge a non-boundary vertex $v' \notin V(F)$ to $u'$, and therefore merge the
    mini-label $v''$ of $v'$ to $u''$ in $P_{i, j}$. Recall that in Case M2 we apply
    the micro-label search operation as a sub-routine to determine the micro label
    $u''$ of $u'$. This routine first searches for $u''$ in the neighborhood of
    $v''$ in $P_{i, j}$, described as Case 1. We first view this case, i.e., the
    edge $\{u'', v''\}$ exists. In this case, the degree of $u''$ is $>0$ before the
    merge by assumption and the degree of $u''$ can become $0$ when $\{u'', v''\}$
    is the only edge incident to $u''$, and the (possibly empty) set of edges
    incident to $v''$ are discarded during the merge to uphold
    Invariant~\ref{inv:nonzero}. Let us now view Case 2, i.e., edge $\{u'', v''\}$ does
    not exists. In the proof of Lemma~\ref{lem:search1} it was shown that in
    Case 2 the edge $\{u, v\}$ (with $u$ and $v$ the global labels of $u'$ and $v'$)
    must be managed by $F$ (as $\{u, v\}$) and therefore Case 2 occurs $O(n/\log n)$
    times over any number of merges in any number of micro graphs $P_{i, j}$. Let us
    assume the worst case, i.e., every time we are in Case 2 the degree of $u''$
    changes in $P_{i, j}$. Each time we must update $\Phi^{>0}_i(u')$. We therefore
    know that at most $O(n/\log^2 n)$ such swaps can occur total. In Case 1 the
    degree can only change from non-zero to zero, but no vice-versa. We can
    therefore bound this by $O(n/\log^2 n + n /\log^2 \log^4 n)$ total. Each update
    takes $O(\log \log^4 n)$ time, and we arrive at a total runtime of $o(n)$.
    \end{proof}
}
\lemminimerge
\full{\proofminimerge}

Contracting edges $\{u, v\}$ in $G$ effectively works exactly as the vertex
merges in mini graphs $P_i$ with the exception
that we do not need to maintain the translation between internal and external
labels, and we do not need to provide free assignment merges for any case. We
again distinguish between three cases: (G1) $u, v \notin \delta G$, (G2) $u \in
\delta G$ and $v \notin \delta G$ and (G3) $u, v \in \delta G$. For Case G2 we
employ a procedure we call \textit{mini-label search procedure}, analogous to
the micro-label search procedure for Case M2.
The following lemma summarizes the edge contraction
operation in $G$.

\newcounter{countlemcontractions}
\newcommand{\lemcontractions}{
  \setcounter{countlemcontractions}{\thetheorem}
\begin{lemma}
    \label{lem:contractions}
    After $O(n)$ initialization time, any number of edge contractions in $G$
    can be computed in $O(n)$ time and uphold the mini-graph responsibility.
\end{lemma}
}
\newcommand{\proofcontractions}{
\begin{proof}
    Contraction of an edge $\{u, v\} \in E$ works mostly analogous to a merge of
    two vertices $u', v' \in V(P_{i})$ with $P_i$ some mini graph. We give a
    quick overview of the three cases, mentioning the differences between the
    analogous cases in mini graphs and how we avoid the forbidden M2 special
    case, outlined in Section~\ref{sec:result}.
    
    \textbf{Case G1.}
    In Case G1 we determine the mini graph $P_i$ that contains (external) mini
    labels of $u$ and $v$, respectively. This is done via $(i, u'):=\phi(u)$ and
    $(i, v'):=\phi(v)$. In $P_i$ we merge $u'$ and $v'$, with the choice of
    which vertex to merge to which being determined by constraints in $P_i$:
    simply put, we have no preferences, and can therefore easily avoid the
    forbidden M2 special case.
    
    \textbf{Case G2.}
    In Case G2 we merge a non-boundary vertex $v \notin \delta G$ with a
    boundary vertex $u \in \delta G$. We therefore require to merge $v$ to $u$
    (Invariant~\ref{inv:4}). We determine the mini graph $P_i$ and the mini
    label $v'$ of $v$ with $(i, v')=:\phi(v)$. We then must determine the mini
    label $u'$ of $u$ in $P_i$ with an operation previously described as
    mini-label search procedure, described in the next paragraph. Assuming we
    have determined $u'$, we then simply merge $v'$ to $u'$ in $P_i$. In the
    case that $v' \in \delta P_i$, this is a free assignment merge. In the case
    that $v' \notin \delta P_i$, we are either free to merge $v'$ to $u'$, or
    are forced to merge $v'$ to $u'$ due to constraints in $P_i$. Thus, we are
    able to avoid the forbidden M2 special case. 
    
    We now discuss the mini-label search procedure needed to determine $u'$.
    Intuitively, this works analogous to the micro-label search procedure, i.e.,
    we simply search in the neighborhood of $v'$ for $u'$. There is a key
    difference to the micro-label search procedure: $u'$ is guaranteed to be in
    $N(v')$ in $P_i$, as the edge $\{u, v\}$ is not contained in $F$ (we are in
    Case G2). To determine $u'$ we iterate over all $x' \in N(v') \cap \Delta
    P_i$, determine $x:=\Phi_i^{-1}(x')$ and test if $x=u$. If this is the case,
    we have found $u':=x'$. We discuss the technical realization of iterating
    over $N(v') \cap \Delta P_i$ in the next paragraph. Under the assumption
    this iteration takes constant time per element the runtime anaylsis works
    exactly as the runtime analysis in the proof of Lemma~\ref{lem:search1}.
    Each edge iterated during this procedure is removed afterwards during the
    merge process (these edges constitute edges that after the merge become
    edges managed by $F$), i.e., each edge is output due to this process only
    once, resulting in a runtime of $O(n)$ total.

    We now describe the technical implementation of this procedure. We need
    additional data structures to implement iteration over $N(v') \cap \Delta
    P_i$ as we need to avoid iterating over neighbors of $v'$ that are not in
    $\Delta P_i$, as otherwise our runtime goal of $O(n)$ can not be upheld.
    First, we distinguish between two cases: (1) $v' \notin \delta P_i$ and (2)
    $v' \in \delta P_i$. In Case 1 we simply determine the micro graph $P_{i,
    j}$ that contains a micro label $v''$ of $v'$ with $(j, v'')=\phi_{i}(v')$.
    We then iterate over all neighbors of $v''$ in $P_{i, j}$ that are colored
    \texttt{boundary} or \texttt{double-boundary}, which are exactly micro
    labels of vertices in $\Delta P_i$. We translate all such neighbors $x''$ to
    mini-labels $x'$ in $P_i$ via $\Phi_{i, j}^{-1}$ or $\phi_{i,j}^{-1}$,
    depending on the color of $x'$. This exactly gives us all neighbors $x' \in
    N(v') \cap \Delta P_i$. For Case 2 we construct additional data structures
    that avoids unecessary iteration over vertices in $N(v') \setminus \Delta
    P_i$. For this we initially (i.e., before any contractions are processed in
    $G$) construct for each $F_i$ a subgraph $F'_i$ that contains all vertices
    of $F_i$, but only edges $\{a', b'\} \in E(F_i)$ with $b' \in \Delta P_i$
    (and $a' \in V(F_i)$). Then, for all $a' \in V(F_i)$ we construct sets
    $\Phi'_i(a') \subseteq \Phi_i(a')$ which contain all tuples $(j, a'') \in
    \Phi_i(a')$ such that $a''$ has neighbors of color \texttt{boundary} or
    \texttt{double-boundary} in $P_{i, j}$. Using these data structures we can
    simply implement the desired mini-label search operation by iterating over
    neighbors of $v'$ in $F'_i$ and then iterating over all $(j, v'') \in
    \Phi'_i(v')$ (for each such $(j, v'')$ execute a search analogous to Case
    1). We maintain $F'_i$ by simply merging two vertices in $F'_i$ exactly when
    we merge two vertices in $F_i$, and adding edges to $F'_i$ exactly when
    edges are added to $F_i$ (and the edges should be contained in $F'_i$). To
    maintain the sets $\Phi'_i$ we construct subgraphs $H'_i$ of $H_i$ that
    maintain exactly these sets. We maintain an analogous invariant to
    Invariant~\ref{inv:nonzero} for the sets $\Phi'_i$, i.e., for all $(j, a'')
    \in \Phi'_i(a')$ for all $a' \in V(F_i)$ it holds that $a''$ has neighbors
    colored \texttt{boundary} or \texttt{double-boundary} in $P_{i, j}$.
    Guaranteeing this modified invariant for all $\Phi'_i$ works exactly as
    Invariant~\ref{inv:2} for all $\Phi^{>0}_i$, shown in the proof of
    Lemma~\ref{lem:minimerge}. In summary, maintaining these additional data
    structures requires $O(n)$ time over any number of edge contractions.

    \textbf{Case G3.}
    In case G3 both $u$ and $v$ are boundary vertices. This case works
    effectively the same as case M3 in mini graphs. We define sets $Z^{v
    \setminus u}$, $Z^{v \setminus u}$ $Z^{u \oplus v}$ analogous to the
    definitions of sets $Z_i^{v^* \setminus u^*}$, $Z_i^{v^* \setminus u^*}$
    $Z_i^{v^* \oplus u^*}$ used in case M3 for merges of two (internal)
    mini-labels $u^*, v^*$ in mini graphs $P_i$. We first iterate over all $(i,
    u', v') \in Z^{u \cap v}$ and merge $v'$ to $u'$ in $P_i$. Note that all
    these merges are free assignment merge in $P_{i}$ as $u', v' \in \Delta P_i$
    for which the mappings $\texttt{internal}$/$\texttt{external}$ are
    available. We then iterate over all $(i, v') \in Z^{v \setminus u}$ and set
    all mappings $\Phi_{i}^{-1}(v'):=u$. Finally, we set $\Phi(u):=Z^{u \oplus
    v}$ and update $\Phi^{>0}(u)$ analogously. The algorithmic realization of
    this works exactly as in case M3 in mini graphs, i.e., the computation of
    these sets (and updates to $\Phi(u)$/$\Phi^{>0}(u)$) is realized via a
    merge of $u$ and $v$ in the graph $H$ and $H^{>0}$. To maintain a runtime of
    $O(n)$ over any number of edge contractions we merge $v$ to $u$ exactly if
    $|\Phi_i(u)| > |\Phi_i(v)|$, analogous to case M3. Clearly any number of
    such edge contractions can be processed in $O(n)$ time in $F$, $H$ and
    $H^{>0}$, and the runtime of merges in all $P_i$ is already stated in
    Lemma~\ref{lem:minimerge}.

    This concludes all three cases, with each case taking $O(n)$ total time for
    contractions. It remains to show how the global graph responsibility is
    upheld. Invariant~\ref{inv:4} is simply upheld by always merging a
    non-boundary vertex to a boundary vertex in case G2, analogous to the same
    invariant regarding mini graphs. To uphold Invariant~\ref{inv:2}.a, recall
    that all merges in mini graphs $P_i$ return a list $L$ of edges that should
    be maintained by $F$ that occurred due to a merge in $P_i$. We simply insert
    these edges to $F$ if they are not yet contained in $F$, and ignore them
    otherwise. The number of edge insertions is $O(n/\log^2 n)$ with each
    insertion taking $O(\log n)$ time. To test if an edge is already contained
    in $F$ we use the $O(1)$ time adjacency query operation provided by
    Lemma~\ref{lem:bvm}. As the runtime of returning all lists $L$ due to merges
    in mini graphs is $O(n)$ time (Lemma~\ref{lem:minimerge}), we arrive at
    $O(n)$ time to maintain Invariant~\ref{inv:2}.a. Maintaining
    Invariant~\ref{inv:nonzero} works analogous to maintaining the same
    invariant in mini graphs: after the contraction of an edge $\{u, v\}$ with
    $u \in \delta G$ and $v \notin \delta G$ (executed by contracting $v'$ to
    $u'$, the respective mini labels of $v'$ and $u'$ in some $P_i$), we simply
    test if $u'$ has neighbors in $P_i$. If this is not the case, we remove $(i,
    u')$ from $\Phi^{>0}(u')$. The runtime analysis works analogous to the
    runtime analysis in mini graphs for maintaining the same invariant, i.e.,
    $O(n)$ time to maintain Invariant~\ref{inv:nonzero}. Invariant~\ref{inv:1}.a
    is upheld by correctly updating the specified mappings during the
    contraction, again analogously to maintaining Invariant~\ref{inv:1}.c in
    mini graphs.

    In summary: any number of edge contractions in $G$ can be processed in
    $O(n)$ time and uphold the global graph responsibility.
\end{proof}
}
\lemcontractions
\full{\proofcontractions}

We additionally provide constant time degree queries. Intuitively, we store the
degree for boundary vertices (in $G$ and each $P_i$) concretely, while for all
other vertices Theorem~\ref{thm:basicenc} provides us with a degree query.

\newcounter{countlemdegree}
\newcommand{\lemdegree}{
  \setcounter{countlemdegree}{\thetheorem}
\begin{lemma}
    \label{lem:degree}
    After $O(n)$ initialization time the degree of any $u \in V$ can be queried
    in constant time.
\end{lemma}
}
\newcommand{\proofdegree}{
\begin{proof}
    To maintain the degree of vertices $u \in \delta G$ store in an array
    $\texttt{Deg}$ the degree of each such $u \in G$. Use an analogous array
    $\texttt{Deg}_i$ for each $P_i$ that stores the degree of $u' \in P_i$. For
    now assume these are correctly maintained. How this is done, is outlined in
    the next paragraph. To query the degree of a vertex $u \in \delta G$ simply
    output $\texttt{Deg}[u]$. To query the degree of a vertex $u \notin \delta
    G$ determine the mini graph $P_i$ that contains the mini label $u'$ of $u$
    via $(i, u'):=\phi(u)$. Note that in this case all incident edges of $u$ are
    incident to $u'$. In $P_i$ query the degree of $u'$ via $\texttt{Deg}_i[u']$
    if $u' \in \delta P_i$, or otherwise determine the micro graph $P_{i, j}$
    that contains the micro label $u''$ of $u'$ via $(j, u''):=\phi_i(u')$ and
    execute the degree query with Theorem~\ref{thm:basicenc}.

    To initialize $\texttt{Deg}$ and each $\texttt{Deg}_i$ use the neighborhood
    iteration (Lemma~\ref{lem:neighborhood}) and simply count the number of
    neighbors. The degree of a vertex only changes if it is involed in a merge,
    i.e., when a vertex $v$ is merged to $u$ (or the analogous merges in mini
    and micro graphs). We only maintain the arrays $\texttt{Deg}$ and
    $\texttt{Deg}_i$ for boundary vertices, all other vertices are automatically
    handles via Theorem~\ref{thm:basicenc}. We now describe how to maintain
    $\texttt{Deg}_i$, maintaining $\texttt{Deg}$ then works analogously.  
    Any time a vertex $v'$ is merged to some $u' \in \Delta P_i$ we distinguish
    between two cases: (1) $v' \notin \delta P_i$ and (2) $v' \in \delta P_i$.
    Recall that in Case 1 we merge $v'$ to $u'$ by merging $v''$ to $u''$ in
    $P_{i, j}$, with $v''$ and $u''$ the respective mini labels of $v'$ and $u'$
    in $P_{i, j}$. To update $\texttt{Deg}_i[u']$ store in $d_0$ the degree of
    $u''$ in $P_{i, j}$ before the merge and in $d_1$ the degree of $u''$ in
    $P_{i, j}$ after the merge. Update
    $\texttt{Deg}_i[u']:=\texttt{Deg}_i[u']+(d_1 - d_0)$. In Case 2 we first
    merge $v'$ to $u'$ in $F_i$, then secondly merge each $v''$ to $u''$ in all
    $P_{i, j}$ that contain both $v''$ and $u''$ (micro labels of $u'$ and $v'$,
    respectively), and thirdly map each $v''$ to be the micro label of $u'$ in
    all $P_{i, j}$ that contain only $v''$, but not $u''$. To update
    $\texttt{Deg}_i[u']$ first set $\texttt{Deg}_i[u']:=\texttt{Deg}_i[u']+(d_1
    - d_0)$, with $d_0, d_1$ the degree of $u'$ in $F$ before and after the
    merge, respectively. Secondly, for each $P_{i, j}$ where $v''$ is merged to
    $u''$, update $\texttt{Deg}_i[u']:=\texttt{Deg}_i[u']+(d_1 - d_0)$ with
    $d_0, d_1$ the degree of $u''$ in $P_{i, j}$ before and after the merge,
    respectively. Thirdly, for $P_{i, j}$ where $v''$ is remapped to be a micro
    label of $u'$, update $\texttt{Deg}_i[u']:=\texttt{Deg}_i[u']+d$ with $d$
    the degree of $v''$ in $P_{i, j}$.

    Each update to the arrays $\texttt{Deg}$ and each $\texttt{Deg}_i$ takes
    constant time for each entry. As we only update an entry in these arrays
    exactly when some other operation is also executed that relates to this
    entry, this is a negligible overhead. The number of bits used to store
    $\texttt{Deg}$ is $O(n/\log^2 n \cdot \log n)=o(n)$ bits and for all
    $\texttt{Deg}_i$ it is $O((n/\log^4 n) \cdot (\log^4 n/ \log^2 \log^4 n) \cdot
    (\log \log n))=o(n)$ bits.
\end{proof}
}
\lemdegree
\full{\proofdegree}

Using the same data structures we use for edge contractions, we can process any
number of vertex deletions in $O(n)$ time. To delete a vertex $u$ we delete all
mini labels $u'$ of $u$ and all micro labels $u''$ of all $u'$. This mostly
works analogously to the contraction algorithm.

\newcounter{countlemvertexdeletion}
\newcommand{\lemvertexdeletion}{
  \setcounter{countlemvertexdeletion}{\thetheorem}
\begin{lemma}
    \label{lem:vertexdeletion}
    Any number of vertex deletions in $G$ can be processed in $O(n)$ time and uphold the
    mini-graph responsibility. %For any $u \in \bar{V}$ we provide a constant
    %time query that answers \texttt{true} exactly if $u \in V$.
\end{lemma}
}
\newcommand{\proofvertexdeletion}{
\begin{proof}
    Let $u \in V$ be a vertex we want to delete. Denote with $u'$ the mini label
    of $u$ in the mini graph $P_{i, j}$ under consideration, and with $u''$ the
    micro label of $u'$ in the micro graph $P_{i, j}$ under consideration. To
    delete $u$ we distinguish between two cases: (1) $u \in \delta G$ and (2) $u
    \notin \delta G$. In Case 1 iterate over all $(i, u') \in \Phi(u)$ and
    delete $u'$ in $P_{i}$ (specified in the next paragraph). Then delete $u$ in
    $F$, $H$ and $H^{>0}$ by deleting all incident edges and adding a boolean
    flag to $u$ that it is marked deleted. In Case 2 delete $u'$ in $P_i$ with
    $(i, u'):=\phi(u)$.

    To delete a vertex $u'$ in a mini graph $P_i$ we distinguish between
    analogous two cases: (1) $u' \in \delta P_i$ and (2) $u' \notin \delta P_i$.
    In Case 1 iterate over all $(j, u'') \in \Phi_i(u')$ and delete $u''$ in
    $P_{i, j}$ (specified in the next paragraph). Then delete $u'$ in $F_i$,
    $H_i$ and $H_i^{>0}$ by deleting all incident edges and by storing a boolean
    flag for $u$ indicating that it is marked deleted. In Case 2 delete $u''$ in
    $P_{i, j}$ with $(j, u''):=\phi_i(u)$. 
    
    All boolean flags mentioned above can be stored in a simple bit vector of
    length $|V(F)|+\sum_i |V(F_i)|=o(n)$. To delete vertices $u''$ in micro
    graphs $P_{i, j}$ use the batch edge deletion of Theorem~\ref{thm:basicenc}
    by deleting all incident edges (in constant time). Afterwards mark $u''$ as
    deleted in $P_{i, j}$, analogous to the same process when merging two
    vertices in $P_{i, j}$ (Section~\ref{sec:table}).

    %To query if a vertex $\bar{u} \in \bar{V}$ is contained in $V$ simply check
    %if it is marked deleted in $F$ (if $u$ is a boundary vertex), or determine
    %the mini graph $P_{i}$ that contains $u'$ via $(i, u'):=\phi(u)$. In $P_i$
    %check if $u'$ is marked deleted in $F_i$ (if $u'$ is a boundary vertex), or
    %determine the micro graph $P_{i, j}$ that contains $u''$ via $(i,
    %u''):=\phi_i(u')$ and check if $u''$ is marked deleted.
\end{proof}
}
\lemvertexdeletion
\full{\proofvertexdeletion}

We are now able to proof Theorem~\ref{thm:main}.
\setcounter{lemma}{0}
\thmmain

\begin{proof}
Construct the dynamic encoding due to Corollary~\ref{lem:initialization}.
Lemma~\ref{lem:neighborhood} gives us the desired neighborhood operation and
Lemma~\ref{lem:degree} the desired degree operation,
Lemma~\ref{lem:contractions} the desired contraction operation and
Lemma~\ref{lem:vertexdeletion} the desired vertex deletions.
\end{proof}

Using hash tables to implement the mappings
$\Phi$, $\Phi^{>0}$, $\Phi_i$ and $\Phi^{>0}_i$ we are able to provide expected
constant time adjacency queries and is able to process any number of edge
deletions in $O(n)$ expected time. Holm et al.\ used the same argument of
replacing a mapping data structure with a hash table to show Lemma~5.15 in their
work~\cite{holm_et_al:LIPIcs:2017:7875}.

\setcounter{lemma}{1}
\coredge

\newcommand{\proofedgedeletions}{
\begin{proof}
Assume $\Phi$, $\Phi^{>0}$, $\Phi_i$ and $\Phi^{>0}_i$ are implemented as hash
tables \cite[Lemma~5.15]{holm_et_al:LIPIcs:2017:7875}, i.e., we can determine
for any $u \in \delta G$ the mini label $u'$ in some $P_i$ (if it exists) in
expected constant time. Let $u, v \in V$ be two vertices for which we want test
adjacency. Denote with $u', v'$ their respective mini labels in some mini graph
$P_i$ and with $u'', v''$ their respective micro labels in some micro graph
$P_{i, j}$. By Invariant~\ref{inv:2}, if the edge $\{u, v\}$ exists, is either managed (1) by $F$, (2) by
some $F_i$ (as edge $\{u', v'\}$) or by $P_{i, j}$ (as edge $\{u'', v''\}$).
Case 1 occurs exactly if both $u$ and $v$ are boundary vertices, i.e., a simple
adjacency query in $F$ suffices to answer the query. In Case 2, one of (or both)
$u$ and $v$ are not contained in $\delta G$. Assume $v \notin \delta G$. We then
determine the mini graph $P_i$ that contains $v'$ and $u'$ via $(i,
v')=\phi_{v}$, we then determine $u'$ either via $\phi(u)$ or $\Phi(u)$ in
expected constant time. If both $u', v' \in \delta P_i$ we can query their
adjacency in $F_i$. Otherwise determine the micro graph $P_{i, j}$ analogous to
determining the mini graph in Case 2, where we then query the adjacency of $u''$ and $v''$. Effectively,
we are able to find in expected constant time the structure ($F$, some $F_i$ or
some $P_{i, j}$) that manages the edge $\{u, v\}$ (with the respective
mini/micro labels of $u$ and $v$ as the endpoints), if it exists.

To delete an edge, first use the adjacency query to find the edge in $F$, $F_i$
or $P_{i, j}$. Subsequently delete the edge there. Deleting all edges in $F$ can
be done in $O((n/\log^2 n) \cdot \log n)=o(n)$ time, deleting all edges in all
$F_i$ can be done in $O((n/\log^4 n) \cdot (\log^4 / \log^2 \log^4 n) \cdot
(\log \log^4 n))=o(n)$ time, and deleting all edges in all $P_{i, j}$ can be done in
$O(n)$ time.
\end{proof}
}
%\full{\coredge}
\full{\proofedgedeletions}

To conclude this section, we show how to modify our encoding so that the labels
remain consistent after modifications. Recall that at the lowest level we
categorize graphs using a lookup table (Section~\ref{sec:table}). The graphs
listed in the lookup table are partially unlabeled, i.e., we only maintain
consistent labeling for boundary vertices. Because of this, after a
modification is executed, e.g., a contraction of some edge $\{u, v\}$ (in the
global graph), some vertices may be re-labeled internally. In detail,
when a contraction is executed at the micro-graph level we effectively switch from
one table entry to antother, i.e., we exchange the index encoding the aformentioned
micro graph. As a consequence, a vertex
labeled with a global label $x$ could now be labeled with a global label $y$.
This is easily mitigated by storing all labeled graphs in the lookup table, but
the encoding is longer succinct. Specifically, this increases the number of
bits required to store an index into the lookup table listing all planar graphs
with $r$ vertices from $\mathcal{H}(r)+o(r)=O(r)$ to $O(r \log r)$
by Corollary~\ref{cor:labeledtable}.
We call an encoding with this change as having a \textit{consisten labeling}.
Our
encoding as previously described uses a lookup table for planar graphs with at
most $r=\log^4\log^4 n=O(\log^4 \log n)$ vertices. 
We encode $O(n/\log^4 n \cdot \log^4 n/r)$ micro graphs with at most $r$ vertices using the
table, and therefore the number
of bits required for the entire encoding increases to $O(n \log \log \log n)$
bits. 
In the next section we show that a consistent 
vertex labeling is
not necessarily required by showing a simple application of testing a graph for
outerplanarity in linear time.
\setcounter{lemma}{21}
\begin{corollary}\label{cor:consistentlabels}
    The encoding of Theorem~\ref{thm:main} can be implemented with consistent
    labels with $O(n \log \log
    \log n)$ bits. Initialization can be done in $O(n)$ time with $O(n \log
    \log \log n)$ bits.
\end{corollary}

\section{Application: Outerplanarity Testing}\label{sec:outer}

We discuss the recognition of outerplanar graphs, which are planar graphs that
can be drawn so that all vertices are incident to the outer face. A well-known
linear-time algorithm exist for recognizing outerplanar exists due to
Wiegers~\cite{Wiegers87}. The algorithm uses $\Theta(n \log n)$ bits of space.
There is also a space-efficient algorithm by Kammer et al.~\cite{KammerKL19},
which runs in $O(n \log \log n)$ time and uses $O(n)$ bits of space. We improve
upon this result by showing how our encoding can be used to implement an
algorithm that runs in $O(n)$ time and bits.

Our approach builds upon the algorithm of Wiegers, which works directly
(ignoring some minor technical details for the moment) with $O(n)$ bits when
using our encoding as a graph data structure, with some minor modifications.
Recall that we use a lookup table at the lowest level of our encoding for
storing all mini graphs with at most $r=\log^4 n \log^4 n$ vertices (as
described in Section~\ref{sec:table}). We extend this lookup table to list
edge-colored planar graphs with a constant number of colors. In detail, we use
the lookup table of Corollary~\ref{cor:coloredtable} that lists all colored
planar graphs with the caveat of a constant factor in the space-usage per
index. In total, this increases the space usage on $O(n/r)$ mini graphs by
$O(r)$ bits each, i.e., an additional $O(n)$ bits total. Thus, the encoding no
longer remains succinct, but as we aim for $O(n)$ bits total, this is
negligible for our use case.

% Also recall that in our encoding, all edges are either managed by micro-graphs (i.e.,
% the lookup table) or by (mini) boundary graphs (as discussed in
% Section~\ref{sec:result}), which store only a few edges. 
% Together with a space-efficient queue data structure this is enough to 
% directly implement Wiegers algorithm using $O(n)$ bits with minimal, such as
% the choice dictionary of Kammer et al.~\cite{KammerS18}.

For convenience, we give a brief sketch of Wieger's algorithm
for a given graph $G=(E, V)$, refer
to~\cite{Wiegers87} for a detailed discussion of corectness. The algorithm
iteratively deletes vertices and contracts edges, coloring the edges either 
as bridge, no bridge, but part of the outer face or as an cross edge part of only inner faces.
Concretely, the set of colors $A=\{\texttt{bridge}, \texttt{out}, \texttt{cross}\}$
is used for the edges. Initially, all edges are colored $\texttt{cross}$.
In the following we describe the set of reduction rules that are applied
iteratively. If a case that is not described is encountered, or
the graph is not reduced to the empty graph, the graph is
not outerplanar. 

\textbf{Case 1.} If a vertex $u$ of degree $1$ exists, delete it. 

The following cases all pertain to a vertex $u$ of degree $2$,
and we denote with $u_1, u_2$ the neighbhors of $u$, with
$e_1=\{u, u_1\}$ ($e_2=\{u, u_2\}$) and with $c_1$ and $c_2$ being
the colors of $e_1$ and $e_2$, respectively. Let $B=\{\texttt{cross}, \texttt{out}\}$.

\textbf{Case 2.1.} If $c_1, c_2 \in B$ and $\{u_1, u_2\} \notin E$,
then contract the edge $e_1$ and set $c_2=\texttt{out}$.

\textbf{Case 2.2.} If at least one of $c_1, c_2$ is equal to $\texttt{bridge}$
and $\{u_1, u_2\} \notin E$, contract $e_1$ and set $c_2=\texttt{bridge}$.

\textbf{Case 2.3.} If $c_1, c_2 \in B$ and $\{u_1, u_2\} \in E$
and $\{u_1, u_2\}$ is colored $\texttt{cross}$,
then contract the edge $e_1$ and set $c_2=\texttt{out}$.

\textbf{Case 2.4.} If $c_1, c_2 \in B$ and $\{u_1, u_2\} \in E$
and $\{u_1, u_2\}$ is colored $\texttt{out}$,
then contract the edge $e_1$ and set $c_2=\texttt{bridge}$.

The entire algorithm is simply implemented by iteratively applying the
specified reduction rules for the outlined cases. Using a queue that maintains
vertices of degree at most $2$ the runtime is linear.

Our space-efficient implementation of this algorithm first encodes the input
graph using our encoding of Theorem~\ref{thm:main}, with the minor
modifications regarding edge coloring outlined at the beginning of the section.
If the construction of the encoding fails, we know that the graph is not
planar, and therefore not outerplanar. Note that the ability to construct the
encoding is not a sufficient criterion for planarity.

The encoding directly allows to perform the necessary edge-contraction
operations and edge-color management. As for a space-efficient implementation
of the queue, we can use the choice dictionary of
Kammer and Sajenko, which uses $O(n)$ bits to maintain a queue over a universe
of $n$ consecutive integers (i.e., the global labels)~\cite{KammerS18}. 
In fact, we can use this queue for boundary vertices directly, since they maintain
a consistent labeling. For the non-boundary vertices 
the
queue requires an additional special implementation: 
For non-boundary vertices, we
maintain a secondary queue that stores the ids of mini-graphs (and at the next
level: micro-graphs) that contain the respective vertices, instead of the
vertex labels itself. For each micro graph the lookup table allows to output in
constant time vertices of degree $1$ or $2$. Altogether, this allows to output
vertices of degree $1$ or $2$ as follows. For boundary vertices in the global
graph, simply use the standard queue. For non-boundary vertices
in the global graph, access the queue that stores ids of mini graphs. We then
repeat analogously in the mini graph given by the id, where we again split
between the cases of boundary vertex and non-boundary vertex. For non-boundary
vertices in the mini-graph we store a queue that stores all degree $1$ or $2$
vertices directly, and for non-boundary vertices we store the ids of micro
graphs that contain vertices of degree $1$ or $2$. At the lowest level we use
the described table lookup technique. Otherwise the algorithm works exactly as
described by Wiegers. We can then easily conclude the following lemma.

\begin{lemma}\label{lem:outerplanar}
    Let $G$ be an unlabaled graph. In $O(n)$ time using $O(n)$ bits we can recognize if $G$
    is outerplanar.
\end{lemma}

\newpage

\bibliography{main}

\end{document}